# Low-cost, high-power mechanical impact transducers for sonar and acoustic through-wall surveillance applications

Franklin Felber[a)]
Physics Division, Starmark, Inc., P. O. Box 270710, San Diego, CA 92198

Abstract

A new concept is presented for mechanical acoustic transmitters and matched resonant receivers. The lightweight, compact, and low-cost transmitters produce high-power acoustic pulses at one or more discrete frequencies with very little input power. The transducer systems are well suited for coupling acoustic pulse energy into dense media, such as walls and water. Applications of the impact transducers are discussed, including detection and tracking of humans through walls and long-duration underwater surveillance by a low-cost network of autonomous, self-recharging, battery-operated sonobuoys. A conceptual design of a sonobuoy surveillance network for harbors and littoral waters is presented. An impact-transmitter and matched-receiver system that detected human motion through thick walls with only rudimentary signal processing is described, and results are presented. Signal processing methods for increasing the signal-to-noise ratio by several tens of dB are discussed.

PACS Numbers: 4338-p, 4320Tb, 4330Jx, 4340Rj

## Sections

---

[a)]Electronic mail: felber@san.rr.com

## 1. Introduction and Background

This paper presents a new concept for mechanical transducers that are particularly well suited for efficiently and inexpensively producing and coupling high-power acoustic pulses into dense media, like walls and water. The transducer system advances reported in this paper and outlined in Fig. 1.1 are:
   (i) a new mechanical transmitter that is compact, lightweight, low-cost, and can operate on battery power, yet produces acoustic pulses in dense media at one or more frequencies that are orders of magnitude more powerful than those produced by alternative transmitters of comparable scale;
   (ii) a resonant receiver matched to the transmitter and designed for efficient coupling to dense media and for enhancing with multiple sensors the signal-to-noise ratio (S/N) of the received signal before signal processing;
   (iii) effective configurations for efficiently coupling the transmitter and matched receiver to dense media; and
   (iv) methods of signal processing that increase S/N by several additional tens of dB.

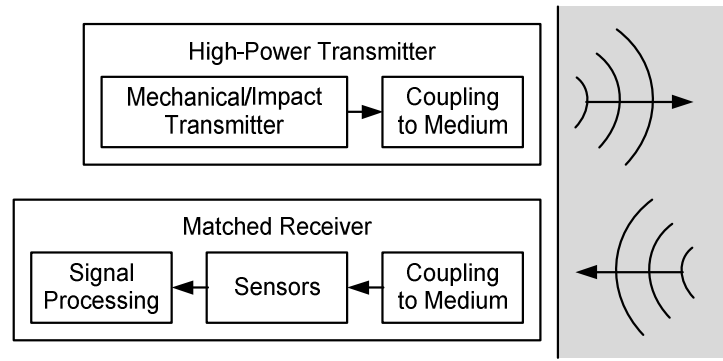

**Figure 1.1.** Overview of novel features of mechanical/impact-transducer system concept presented in this paper.

Sections 2 through 4 present the innovative concepts of mechanical impact transmitters and matched resonant receivers and their effective coupling into solid and liquid media. Sections 5 through 9 focus on two of the applications that make good use of the unique features of the new transducer system: (i) Detecting and tracking humans through walls, particularly the steel walls of cargo containers and trailer trucks, which cannot be penetrated by radar; and (ii) low-cost, lightweight, battery-operated sonobuoys proliferated in networks that can autonomously perform underwater surveillance for years in harbors and littoral waters.

An acoustic means of through-wall surveillance (TWS) and tracking was discovered by the author in 1997. While conducting programs to develop acoustic sensors of concealed weapons [1 – 6], it was found that commercial off-the-shelf (COTS) ultrasound transducers placed against a solid barrier produced an echo from the other side of the barrier that changed when someone moved behind the barrier. Then by subtracting successive echo pulse waveforms, the difference waveform, through destructive interference, revealed only those persons or objects that moved between pulses; the echo pulse waveforms returned from stationary objects canceled each other. The round-trip time of each pulse returned to a receiver from a moving person or object indicated its range, and triangulating the ranges to multiple receivers indicated its location.

The advantages of acoustic TWS, and particularly handheld and portable systems, over alternative TWS technologies quickly became apparent. The primary alternatives to acoustic TWS since the mid-1990s have been radar-microwave and passive millimeter-wave sensors [7 – 20]. Other sensor technologies, like infrared sensors, could image weapons concealed on a body beneath clothing, but could not image through walls. Passive millimeter-wave sensors required illumination of the targets by millimeter-wave radiation from the sky, which effectively limited their applicability to finding persons in areas open to the sky [11, 14]. Radar-microwave sensors, like the Hughes Motion Detection Radar [10, 12, 16], differential radar [13], radar 'flashlight' [15], Time Domain's RadarVision and SoldierVision [17 – 19], and Livermore's Urban Eyes [20], were limited by attenuation in walls to long microwave wavelengths, typically S band and longer, which did not allow detection of the millimeter-scale motions of 'motionless' people. More significantly, radar could not penetrate metal or metal-lined walls or even the aluminum-backed fiberglass insulation typically found in homes and buildings.

SAIC's Vehicle and Cargo Inspection System (VACIS®) [21] is designed to penetrate 15 cm of steel and image the entire contents of cargo containers and trailer trucks, but the system elements must be big enough to span the trucks and cargo containers it scans. Also, highly ionizing gamma radiation is not allowed for use on humans in TWS applications, which is unfortunate, because it images humans through steel walls very well (see, *e. g.*, [22]).

The foremost advantage of acoustic TWS is that sound penetrates metal walls almost as well as other high-impedance wall materials, and does so with harmless non-ionizing radiation. Another significant advantage is that acoustic TWS is sensitive to motions smaller than 1/10 of a wavelength. Because the signal processing destructively interferes successive waveforms with each other, a movement on the scale of one wavelength will produce an interference waveform comparable in signal strength to the successive echo waveforms themselves. But even a movement on a scale less than about 1/10 of a wavelength can produce interference waveforms having a good fraction of the signal strength of the echoes.

Following the discovery of acoustic TWS, we developed an acoustic TWS breadboard system [23], shown schematically in Fig. 1.2(a). The system operated by transmitting narrowband acoustic pulses through a wall and receiving the reflected signals from all objects inside the room at two laterally separated receivers. Signal processing algorithms, presented in Sec. 6, were used to track multiple persons through a wall and to detect even the millimeter-scale chest motion of a person lying down and breathing quietly. Figure 1.2(b) shows an implementation of our tracking algorithm on TWS binaural waveforms from the breadboard TWS monitor of Fig. 1.2(a).

Although the COTS ceramic piezoelectric transducers used in this system were operated near their damage threshold, 3200 volts peak-to-peak, TWS performance was severely power-limited, especially since *these transmitters lose their resonant qualities when placed against a wall*. When placed against a wall, their quality factor $Q$ is reduced from 25 by more than an order of magnitude, and the acoustic power, which scales as $Q^2$, falls by a factor of hundreds. Not even COTS underwater transducers [24] could provide the high-power, narrowband acoustic pulses needed for TWS. As seen in Fig. 1.2(b), the S/N was marginal at a range less than 3 m through a few-cm-thick wooden door.

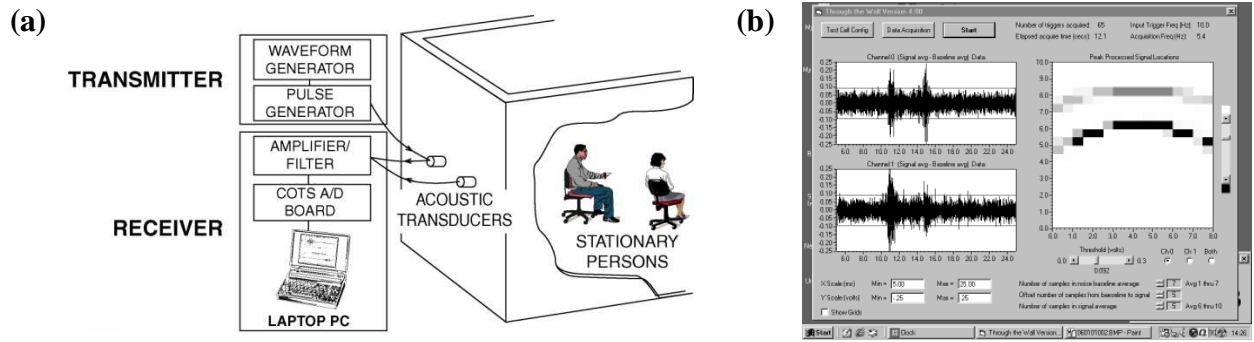

**Figure 1.2. (a) Schematic of breadboard acoustic TWS monitor. (b) Monitor shows two stationary persons sitting 6 ft and 8 ft behind a 4.3-cm solid wooden door.**

In response to the need for high-power TWS transducers, we developed a narrowband, tunable mechanical transmitter for the Marine Corps Systems Command (MCSC) that was capable of scaling up the transmit / receive power of the acoustic TWS system by more than 72 dB (a factor of 16 million), but that was lighter than the COTS transmitter it replaced [25]. Another significant advantage of this mechanical "MCSC transmitter", shown in Fig. 1.3, was safety from high voltages, because it operated on a single 9-V battery.

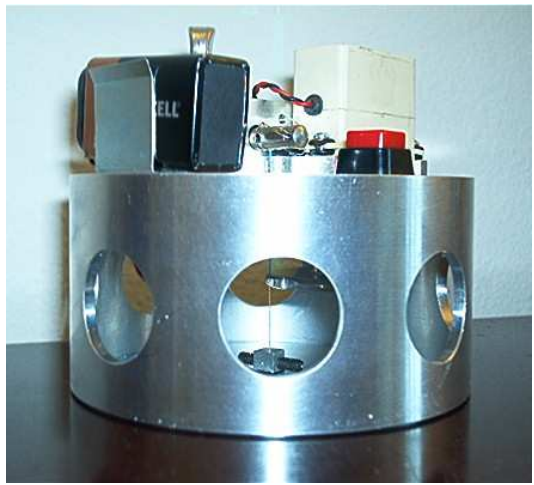

**Figure 1.3. Breadboard transmitter developed for MCSC TWS program [25].**

This tunable narrowband transmitter is based on the following concept. A flat surface is caused to vibrate at high acoustic frequency ($\geq 7$ kHz) by battery-operated mechanical means, rather than by high voltages. In the embodiment shown in Fig. 1.3, the transmitter comprises a thin plate, a wire under tension, a mechanical means of plucking the wire, and a supporting structure. The vibrating surface is a thin, circular aluminum plate, held tightly against a lip on the bottom of the structure in Fig. 1.3 by the tension of the wire. When the wire is plucked, it causes the plate to oscillate longitudinally at twice the frequency of transverse wave oscillations on the wire, which is much higher than the natural frequency of the plate. The thin plate effectively acts as a spring, with the stiffness of the plate providing the restoring force.

This design had the advantage of tunable frequency (by adjusting the tension in the wire), but the disadvantage of limited amplitude. The transmitter designs in Sec. 2 of this paper have an important modification of the MCSC transmitter design that boosts the peak power output by orders of magnitude. In the MCSC transmitter, after the wire is plucked, the amplitude of its oscillations is very much smaller than the initial amplitude. Since power scales as the square of the plate displacement, a nontunable, resonant configuration, such as is shown in Fig. 1.4, has orders of magnitude greater power, as is demonstrated in Sec. 5.

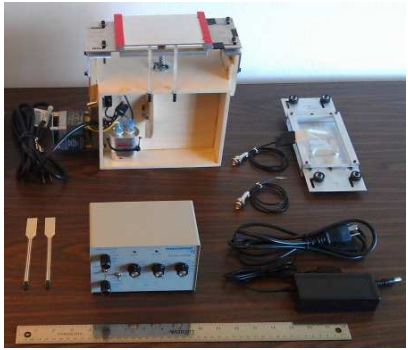

**Figure 1.4. Complete breadboard TWS transducer system, size relative to 18-in ruler.**

Figure 1.4 shows the breadboard TWS resonant transducer system, which is described in Sec. 8. From left to right and back to front, the items in Fig. 1.4 are: Impact transmitter; resonant receiver plate assembly; triggering cable; 2 piezo-film sensors; and a COTS filter/amplifier with transformer and power cord.

Developing high-power transducers is only part of the challenge for acoustic TWS. The transducers must not only be designed for efficient coupling of their acoustic energy to solid walls, but the design of the transducer/wall interface itself is of critical importance to overall coupling efficiency. Indeed, others who have tried to develop acoustic TWS systems to compensate for the shortcomings of radar have found that "the interface between the acoustic transducers and the surface of the wall" was a challenge that "needed to be overcome" [19]. Section 4 discloses an effective way to couple mechanical transducers to walls.

The high-power transmitter concept disclosed in Sec. 2 was developed for underwater and through-wall applications. But since the issues concerning design of transducers and coupling to water and walls are related to those issues for all dense media, the TWS transducer technology presented in Secs. 3 and 4 is applicable, with appropriate modifications, to production and coupling of high-power acoustic pulses to all liquid and solid media as well. The coupling into liquid and solid media is discussed in Sec. 4, and a conceptual design of a sonobuoy surveillance network using these technologies is presented in Sec. 9. The advantages of the mechanical transducers described in this paper – high acoustic pulse power, low electrical power, high efficiency, light weight, compact size, low cost, low-voltage long-duration battery operation – are as attractive for underwater systems as for TWS.

Section 9 presents a conceptual design of a *complete autonomous sonobuoy system*, including a 1-kHz impact-transducer active sonar, processing, power, telemetry, and flotation, that radiates intermittent pulses at 191 dB *re* 1 µPa at 1 m under battery power at a few volts, and that should cost less than or about $1000, when manufactured in quantity.

In short, COTS transducers and power amplifiers, even those that are inefficient and have relatively less demanding electrical power requirements, are quite expensive. The power conditioning subsystems generally are heavy and bulky and require the order of kilovolt voltages. None approaches a cost of hundreds of dollars per unit, which is a reasonable goal for battery-operated, amplifier-free, high-power impact transducers.

Following the summary and conclusions in Sec. 10, the abstract of the U.S. patent application based on the material in this paper is published in Appendix A.

## 2. Concept of Mechanical Impact Transmitters

In our program to develop a high-power acoustic transmitter for the MCSC [25], we discovered a highly effective and efficient way to couple acoustic energy into a dense medium, such as a wall or water: *A properly designed thin plate transduces a <u>mechanical impulse</u> to acoustic energy in a dense medium with high efficiency at the resonant mode frequencies of the plate.*

The concept for transduction by impact excitation of a thin plate is shown in Fig. 2.1(a). An electromechanical actuator propels a mass, called an impactor, at a thin plate. The mass impacts the plate and bounces back, exciting the fundamental mode if striking the center of a symmetrical plate, and exciting higher frequency modes if striking off center. The thin plate then rings down, delivering much of its kinetic energy to acoustic radiation in the medium, if coupled properly.

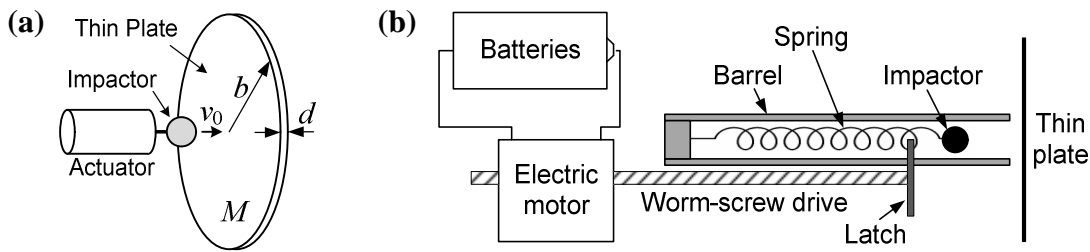

**Figure 2.1. Conceptual sketches of: (a) impact transmitter; and (b) electromechanical pull-type actuator.**

In a sense, this 'impact transmitter' performs much like a spring-loaded mousetrap. Potential energy is accumulated and stored at low power over a relatively long time. Then by opening a latch or some other mechanical release mechanism, the potential energy is suddenly transformed to kinetic energy and delivered to a target in an instant at high power.

Although impact transmitters can have high energy efficiency, that is not their principal advantage or what will allow the transmitters to be manufactured at low cost. *The principal advantage of impact transducers is that the energy for each pulse is accumulated over long times at low powers, and therefore can be operated with ordinary batteries and no power conditioning.* For example, the TWS transmitter in Fig. 1.3, driven by a small electric motor powered by one 9-volt transistor battery, produced orders of magnitude higher acoustic power than the breadboard TWS system in Fig. 1.2, driven by thousands of volts peak-to-peak.

To understand how the impact transmitter works in practice, consider the sketch of the pull-type actuator in Fig. 2.1(b). This actuator type is not the most efficient, but it may be the simplest conceptually because of its resemblance to a mousetrap. A small, battery-operated electric motor powers a worm-screw drive, which pulls back a spring. The compressed spring is held in place by a latch. When the latch is released, the spring drives the impactor, which can be a steel sphere, just past the equilibrium length of the spring, to strike the thin plate.

A more common and more efficient pull-type actuator, though not as simple conceptually, is one that uses a solenoid and permanent magnet in place of the electric motor and worm-screw drive in Fig. 2.1(b). In all pull-type actuators involving a spring, the impactor should strike the plate just beyond the equilibrium length of the spring, so that the full potential energy of the spring is converted to kinetic energy, and the impactor strikes the plate before decelerating against the spring's pull.

Figure 2.2 shows several means of mechanically driving a thin plate, which is part of the interface of the transmitter with the dense medium. Each employs efficient electric motors and/or actuators with decades of development and commercial and industrial use.

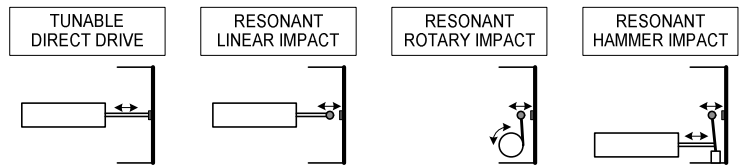

**Figure 2.2. Mechanical means of driving a thin-plate transmitter with motors and/or actuators.**

The direct-drive mechanism is familiar from uses in high-power hydraulic transmitters and is a common means of driving loudspeakers. The MCSC transmitter shown in Fig. 1.3 most resembles a tunable direct drive mechanical actuator. The principal disadvantage of direct drive in this context is that the oscillating surface must be driven at high power to produce high-power radiation. It produces a displacement of the plate limited to the stroke of the motor. The other three mechanisms shown in Fig. 2.2 are all impact-type transmitters. The pull-type actuator in Fig. 2.1(b) is an example of a linear-impact mechanical actuator. The rotary-impact and hammer-impact transmitters employ mechanical advantage.

In the following we present general design equations for an impact transmitter. For specificity, a circular plate with edge-clamped boundary conditions is chosen, but the design equations are easily modified to other plate shapes and boundary conditions. Then a very specific point design of an underwater impact transmitter is presented.

The normal displacement $z$ at radius $r$ and time $t$ of a thin circular plate, rigidly clamped all around its circumference, vibrating in its fundamental mode, is [28]

$$z(r,t) = z_0(t)\sin(\omega t)[J_0(3.2r/b) + 0.0555 I_0(3.2r/b)]/1.0555. \tag{2.1}$$

Here, $z_0(t)$ is the displacement amplitude at the center of the plate, which decays with time; $J_0$ and $I_0$ are the zero-order Bessel function and modified Bessel function of the first kind, respectively; $b$ is the radius of the plate (to the clamped edge); and $\omega$ is the angular frequency of oscillation. The frequency of the fundamental mode of the thin plate in air is [28]

$$f_p = 0.47(d/b^2)[Y/\rho_p(1-\sigma^2)]^{1/2}, \tag{2.2}$$

where $d$ is the thickness, $\rho_p$ is the density, $Y$ is the Young's modulus, and $\sigma$ is the Poisson's ratio of the plate material. To a fair approximation, the fundamental frequency of the plate is $d/A$ times the speed of sound in the plate, where $A = \pi b^2$ is the area of the plate.

Unlike a piston transmitter [see, *e.g.*, 24], the impact transmitter depends on the stiffness of the plate to provide the restoring force. That means the surface of the vibrating plate is elastically deformed into shapes that are determined by the modes of oscillation and by the boundary conditions. An edge-clamped circular plate oscillating in its fundamental mode, for example, is elastically deformed into the shape given by Eq. (2.1).

When the surface is deformed, only some portion of the surface effectively contributes to coupling acoustic power and energy into the medium. The central region of an edge-clamped circular plate vibrating in its fundamental mode radiates much more acoustic power than the edge region does, because the amplitude is greater at the center than near the edge. The following calculations, therefore, are based on an *active-area-piston approximation*. That is, the complicated plate shape of Eq. (2.1) is modeled by a flat circular piston having the same velocity as the on-axis velocity of the edge-clamped plate, and the same total energy, but a smaller 'active' area $A_a = \pi a^2$ and a smaller 'active' radius $a$. The plate mass is $M = \rho_p A d$, but the 'active' plate mass is only $M_a = \rho_p A_a d = (a/b)^2 M$. This model is justified because only an 'active' portion of the plate near the center effectively contributes to acoustic power.

In the active-area-piston approximation then, the edge-clamped plate is represented by a baffled flat piston of area $A_a$ and mass $M_a$ with a displacement given by $z_a(t) = z_0(t)\sin(\omega t)$ and a velocity given by $u_a(t) = u_0(t)\cos(\omega t)$, for $|d\ln(u_0)/dt| \ll \omega$. The total (kinetic plus potential) specific energy of the piston is $E(t) = u_0^2(t)/2$.

From Eq. (2.1) and its first derivative with respect to time, the maximum kinetic energy of the whole plate is calculated as a function of the maximum plate velocity at the center, $u_0$, and is found to be $0.182(Mu_0^2/2)$, which is equal to the maximum kinetic energy of a flat circular piston having the same velocity, $u_0$, but a mass of only $0.182M$. Because only the central region of the edge-clamped plate moves much, the active plate mass $M_a$ is only 18.2% of the actual mass. That means the active area of the plate is $A_a = 0.182A$, and the active radius is $a = 0.427b$. Different boundary conditions would result in different active masses and areas.

When the impact transmitter is submerged in water or other dense fluid, the fluid increases the effective mass of the plate, decreases the resonant frequency, increases the maximum acoustic power output, and broadens the bandwidth, compared to the performance of the same transmitter in air. In the low-frequency limit, $ka \ll 1$, where $k = \omega/c$ is the wavenumber and $c$ is the speed of sound in the fluid, the low-frequency radiation resistance and radiation reactance of a cylindrical baffled piston of area $A_a$ are, respectively [24, 28],

$$R_r = ZA_a(ka)^2/2, \qquad (2.3)$$

$$X_r = ZA_a(8ka/3\pi), \qquad (2.4)$$

where the characteristic acoustic impedance $Z = \rho c$ is the product of the fluid density $\rho$ and $c$.

Since $X_r \gg R_r$ in the low-frequency limit, the radiation reactance is like that of a mass [28],

$$M_r = X_r/\omega = (8a^3/3)\rho. \qquad (2.5)$$

That is, the low-frequency reactance affects the plate oscillations just as would adding to the active mass of the plate an additional mass $M_r$. The effective mass of the oscillator becomes $M_e = M_a + M_r$, which downshifts the angular frequency of the fundamental plate resonance in the fluid from $2\pi f_p$ in air to

$$\omega = 2\pi f_p/\mu, \qquad (2.6)$$

where $\mu \equiv (M_e/M_a)^{1/2}$.

The power radiated into the fluid by the transmitter, averaged over a wave cycle, is related to the radiation resistance by [28]

$$P(t) = R_r E(t) = ZA_a(ka)^2 E(t)/2. \qquad (2.7)$$

Since the power that is radiated into the fluid is equal to the rate of loss of the total effective energy $M_e E$, the total specific energy decays exponentially as

$$E(t) = E_0 \exp(-\Gamma t), \qquad (2.8)$$

where $\Gamma \equiv R_r/M_e$ is the decay rate, and $E_0$ is the initial total specific energy of the plate after the impactor has imparted its energy.

The quality factor of the oscillator is $\omega / P(t)$ times the effective energy $M_e E(t)$. But since $P(t) = R_r E(t)$, the quality factor is

$$Q = \omega / \Gamma . \tag{2.9}$$

By definition [24], the transmitter bandwidth is $B = \omega / 2\pi Q$, or

$$B = \Gamma / 2\pi . \tag{2.10}$$

The potential energy of the effective mass is of the form $K(z_0 \sin \omega_0 t)^2 / 2$, where the effective spring constant of the submerged plate is

$$K = M_e \omega^2 . \tag{2.11}$$

Combining Eqs. (2.6) and (2.11) shows that the effective spring constant of the plate,

$$K = M_a (2\pi f_p)^2 , \tag{2.12}$$

is the same, whether the plate is in air or submerged in fluid. The potential energy is a maximum when the plate reaches its maximum displacement amplitude, $z_0(t)$, at which point the potential energy $K z_0^2(t) / 2$ equals the total energy $M_e E(t)$. The displacement of the plate on axis, therefore, is

$$z(t) = (2 E_0 / \omega^2)^{1/2} \sin(\omega t) \exp(-\Gamma t / 2) , \tag{2.13}$$

and the velocity of the plate on axis, for $|d \ln(u) / dt| \ll \omega$, is

$$u(t) = (2 E_0)^{1/2} \cos(\omega t) \exp(-\Gamma t / 2) . \tag{2.14}$$

Since an efficient transmitter uses a lightweight plate to move lots of fluid mass at high frequency, a transmitter plate material with low density and high sound speed is preferred. For specificity, the general design equations above are now presumed to apply to an aluminum plate, which has the following properties [28]: $\rho_p = 2700 \text{ kg/m}^3$, $Y = 7.1 \times 10^{10} \text{ Pa}$, $\sigma = 0.33$. Also for specificity, the transmitter is considered to be operating in sea water, which has the following properties at $13°C$ [28]: $\rho = 1026 \text{ kg/m}^3$, $c = 1500 \text{ m/s}$, $Z = 1.54 \times 10^6 \text{ kg/m}^2\text{s}$.

When the general design equations above are applied to a circular, edge-clamped aluminum plate operating in sea water, the only unspecified parameters are the plate radius $b$, plate thickness $d$, and specific energy $E$. In terms of these three parameters, the design equations then become

a) Plate frequency in air  $f_p = (260\,\text{kHz}\cdot\text{cm})\,d/b^2$

b) Plate area  $A = \pi b^2$

c) Active area  $A_a = 0.57 b^2$

d) Active radius  $a = 0.43 b$

e) Plate mass  $M = (8.5\,\text{g}/\text{cm}^3)\,b^2 d$

f) Active mass  $M_a = (1.5\,\text{g}/\text{cm}^3)\,b^2 d$

g) Reactive mass  $M_r = (0.21\,\text{g}/\text{cm}^3)\,b^3$

h) Mass ratio, $M_e / M_a$  $\mu^2 = 1 + 0.14\,b/d$

i) Effective mass  $M_e = (1.5\,\text{g}/\text{cm}^3)\,\mu^2 b^2 d$

j) Effective energy  $M_e E = (1.5\,\text{g}/\text{cm}^3)\,E\mu^2 b^2 d$

k) Angular frequency  $\omega = (1600\,\text{kHz}\cdot\text{cm})\,d/\mu b^2$           (2.15)

l) Wavenumber  $k = 11\,d/\mu b^2$

m) Diffraction constant  $ka = 4.6\,d/\mu b \ll 1$

n) Spring constant  $K = (4.0\times10^9\,\text{kg}/\text{s}^2\text{cm})\,d^3/b^2$

o) Max. displacement  $z_0 = (8.8\times10^{-7}\,\text{s}/\text{cm})\,E_0^{1/2}\,\mu b^2/d$

p) Max. velocity  $u_0 = 1.4\,E_0^{1/2}$

q) Radiation resistance  $R_r = (920\,\text{kHz}\cdot\text{g}/\text{cm}^2)\,d^2/\mu^2$

r) Radiation reactance  $X_r = (340\,\text{kHz}\cdot\text{g}/\text{cm}^2)\,bd/\mu$

s) Radiated power  $P = (920\,\text{kHz}\cdot\text{g}/\text{cm}^2)\,Ed^2/\mu^2$

t) Decay rate  $\Gamma = (590\,\text{kHz}\cdot\text{cm})\,d/\mu^4 b^2$

u) Bandwidth  $B = (95\,\text{kHz}\cdot\text{cm})\,d/\mu^4 b^2$

v) Quality factor  $Q = 2.70\,\mu^3$

For a demonstration of the use of Eqs. (2.15) in the point design of an undersea impact transmitter, consider the following sample specification:

*An edge-clamped circular aluminum plate is to produce acoustic pulses in sea water at a central frequency of 1 kHz and with a bandwidth broad enough that 90% of the plate energy is to be radiated in the first 4 ms, that is, during the first four wave cycles, at an average power of 50 W.*

This specification imposes three conditions on the three unspecified parameters, $b$, $d$, and $E_0$. (The mass ratio $\mu^2$ in Eq. (2.15h) is a function of $b$ and $d$.) From Eq. (2.8), an energy decay of 90% in 4 ms implies $\Gamma = 580\,\text{Hz}$ and $Q = 11$. Then Eq. (2.15v) gives $\mu^2 = 2.5$, Eq. (2.15h) gives $b = 11d$, and Eq. (2.16k) gives the plate radius and thickness as $b = 14\,\text{cm}$ and $d = 1.3\,\text{cm}$. Since the average power radiated in the first 4 ms is 50 W, so that 200 mJ of radiated energy is 90% of the initial energy, the initial specific energy must be $E_0 = (220\,\text{mJ})/M_e$, times an efficiency factor to account for conversion of plate energy to acoustic energy. From Eq. (2.15i), the effective mass of the plate (plus sea water) is $M_e = 1.0 \, \text{kg}$. The impactor kinetic energy, therefore, must be 210 mJ, times a factor to account for efficiency of conversion of impactor kinetic energy to acoustic energy. The efficiency is optimized for an impactor mass about equal to $M_e$.

A preliminary point design of the impact-transmitter plate for this example is given in Table 2.1. The values in the table are consistent with all the design equations, Eqs. (2.15), and with the specifications that the transmitter radiate 50 W average power at 1 kHz for 4 ping-ms. The overall accuracy of the estimates in the table is at best of the order of 10%. The two largest sources of error are the *active-area-piston approximation* discussed above and the neglected terms of order $(ka)^2 \sim 7\%$.

**Table 2.1. Impact transmitter point design for edge-clamped aluminum plate that can radiate 50-W, 4-ms pulses underwater at 1 kHz.**

| Parameter | Value |
|---|---|
| Plate radius, cm | $b = 14$ |
| Plate thickness, cm | $d = 1.3$ |
| Initial specific energy, J/kg | $E_0 = 0.21$ |
| Plate frequency in air, kHz | $f_p = 1.6$ |
| Plate area, cm$^2$ | $A = 650$ |
| Active area, cm$^2$ | $A_a = 120$ |
| Active radius, cm | $a = 6.1$ |
| Plate mass, kg | $M = 2.2$ |
| Active mass, kg | $M_a = 0.41$ |
| Reactive mass, kg | $M_r = 0.63$ |
| Effective mass, kg | $M_e = 1.0$ |
| Mass ratio | $\mu^2 = 2.5$ |
| Initial effective energy, J | $M_e E_0 = 0.22$ |
| Angular frequency, kHz | $\omega = 6.3$ |
| Wavenumber, cm$^{-1}$ | $k = 0.042$ |
| Diffraction constant | $ka = 0.26$ |
| Spring constant, kg/(ms)$^2$ | $K = 41$ |
| Max. displacement, cm | $z_0 = 0.010$ |
| Max. velocity, cm/s | $u_0 = 65$ |
| Radiation resistance, kg/ms | $R_r = 0.60$ |
| Radiation reactance, kg/ms | $X_r = 3.9$ |
| Peak radiated power, W | $P_0 = 260$ |
| Average radiated power, W | $P_r = 50$ |
| Decay rate, kHz | $\Gamma = 0.58$ |
| Bandwidth, kHz | $B = 0.091$ |
| Quality factor | $Q = 11$ |

The acoustic power radiated during each pulse,

$$P(t) = P_0 \exp(-\Gamma t) \cos^2(\omega t), \tag{2.16}$$

and averaged over a wave cycle, $(P_0/2)\exp(-\Gamma t)$, is shown in Fig. 2.3.

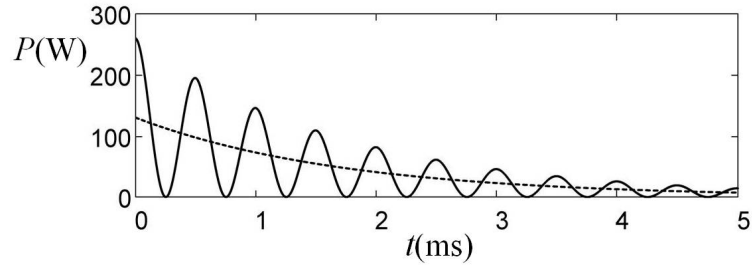

**Fig. 2.3.** Radiated power *vs*. time (solid curve) for point design of 1-kHz undersea impact transducer; averaged over wave cycle (dashed).

The optimal impactor for the plate of Table 2.1 is found by trading off limitations imposed by optimizing efficiency with respect to: (i) frequency; (ii) power; and (iii) duration of impact. The choice of impactor material is governed by the need to have the impactor rebound from the plate quickly. The duration of the impact on the plate should be a small fraction of the oscillation period of the plate, in order that the impactor not impede the first oscillation of the plate after impact. Small, hard impactors produce impacts of short duration. The impact duration is proportional to the diameter of a spherical impactor, and hard impactors produce more nearly elastic collisions. A good choice of impactor material, therefore, is steel.

The most efficient energy transfer from a steel sphere to an aluminum plate occurs when the mass of the sphere $m$ is about equal to the effective mass of the plate (plus sea water), $M_e = 1.0$ kg in this example, corresponding to a diameter of the steel sphere of $D = 6.4$ cm, or about 2.5 in. Ideally, the impact is impulsive, with a duration $\Delta t$ lasting much less than a wave period, that is, $\omega \Delta t \ll 2\pi$, so that the impactor will not interfere with the free oscillation of the plate. Then the impactor mass $m$ can be matched to the effective mass $M_e$ for optimal efficiency of energy transfer. If the impact duration is too long, however, then the impactor must deliver the necessary energy to the plate at higher velocity and lower mass and lower efficiency. The energy efficiency of elastic impacts on a plate falls with impactor mass as

$$\varepsilon = 4(m/M_e)/(1+m/M_e)^2, \tag{2.17}$$

as shown in Fig. 2.4.

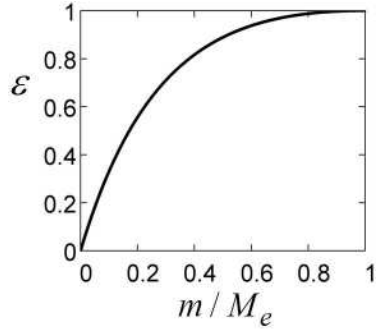

**Figure 2.4. Efficiency of energy transfer for elastic impacts of impactor mass $m$ on plate of effective mass $M_e$.**

The acoustic properties of steel spheres impacting hard surfaces have been studied since the 1970s within the field of civil engineering known as impact echo. Impact-echo systems are used for nondestructive testing, mainly of concrete and masonry structures. A short impact, produced by striking a small steel ball against a surface, generates sound waves in the structure that are reflected strongly at surfaces and at flaws.

The impact duration, $\Delta t(\mu s) = 43 D(\text{cm})/h^{0.1}(\text{m})$, of a steel sphere dropped onto a concrete slab [29] is directly proportional to the diameter $D$ of the sphere and nearly independent of drop height $h$. In terms of impact speed $v$, the duration of impact is

$$\Delta t = (43\,\mu s) D(\text{cm})/[v/(4.4 \text{ m/s})]^{0.2} . \tag{2.18}$$

For a 1.0-kg steel sphere to radiate the specified acoustic pulses, its kinetic energy at impact must be 0.22 J times an efficiency factor to account for conversion of plate energy to acoustic energy. Naively applying Eq. (2.18), the impact duration for steel on concrete, to impacts of steel on aluminum, suggests the impact of a 1-kg steel sphere at a speed of 0.65 cm/s lasts 0.4 ms, which is a good fraction of the 1-ms wave period.

Examining the detailed dynamics of the impacts of steel spheres on aluminum surfaces is beyond the scope of this paper. If it is necessary to use an impactor of higher velocity and lower mass, the energy transfer efficiency could be affected. It should be possible, however, to find some combination of plate design parameters and impactor mass and speed that still meets specifications and has an impactor-to-plate energy transfer efficiency of at least about 50%. Even if the mechanical energy of the impact transmitter in this point design is accumulated as potential energy over just a few seconds between pulses, the battery power to a 50-W impact transmitter might well be less than 1 W.

As was indicated in Figs. 2.1 and 2.2, an impact transmitter requires a mechanical device or actuator to cause the impactor to be repeatedly accelerated against the plate at the pulse repetition frequency of the transmitter. For the point design above, the actuator should be capable of accelerating a steel sphere to an energy of at least about 1 J, assuming an impactor-to-acoustic-energy transfer efficiency of only 25%. To keep the transducer design compact, the acceleration distance should be not much more than about 5 cm, which requires a spring constant of the actuator spring of 800 kg/s$^2$. The force on the latch holding the actuator spring at full compression is 40 N, or 9 lbs. Assuming one pulse every 5 seconds for the underwater surveillance application in this example, the average power required for a worm-screw drive, or any similar driver, to compress the actuator spring by 5 cm in 5 s is 0.2 W. A variable-pitch worm-screw drive can be used to level the operating power. If the battery to mechanical power efficiency is 30%, then the average electrical battery power is 0.7 W. The design parameters of the actuator spring are summarized in Table 2.2.

**Table 2.2.   Nominal point design of actuator spring.**

| Parameter | Value |
|---|---|
| Spring constant | 800 kg/s$^2$ |
| Compression distance | 5 cm |
| Force at max. compression | 40 N (9 lbs.) |
| Time to compress | 5 s |
| Mech. power to compress | 0.2 W avg. |

The estimated energies and efficiencies for impact transduction with the point design are summarized in Table 2.3. The overall efficiency for radiating 50 W for 4 ping-ms is estimated to be 7.5%. Table 2.3 shows that a battery energy/pulse of 3.3 J was needed for an impact transducer to radiate a 1-kHz, 4-ping-ms pulse at an average power of 50W. Perhaps some transmitters can be designed to radiate the same power and energy with comparable electrical-to-acoustic efficiency, but no others operate at less than about 1 W of battery power or can be built so inexpensively.

**Table 2.3.   Summary of estimated energies and efficiencies for point design.**

| Parameter | Efficiency | Energy (J) |
|---|---|---|
| Battery energy/pulse |  | 3.3 |
| Battery to stored mech. | 30% | 1 |
| Stored mech. to plate KE | 50% | 0.5 |
| Plate KE to acoustic pulse | 50% | 0.25 |
| TOTAL | 7.5% |  |

### 3.  Concept of Matched Resonant Receivers

As a general rule, virtually any type of acoustic transmitter, with appropriate modifications, can also be operated as a receiver. The impact transmitter is no exception. The thin plate that is the active surface in an impact transmitter can also be used to receive returning echo signals in TWS or sonar applications.

The receiver plate can be the same plate used for transmission, but it does not have to be. For many applications, it may be preferable to have the receiver plate be separate from the transmit-ter plate, even if both plates are collocated. A common reason for wanting to have separate transmitters and receivers is that the echo signal is expected to return to the receiver before the oscillations of the transmitter plate have decayed sufficiently. In such cases, the transmitter and receiver plates might not only need to be separate, but also acoustically isolated from each other. For related reasons, the receiver might also need to be time-gated, so that it does not start 'listening' for returning echo signals until some number of milliseconds after the transmitter generates a pulse. For the reasons that follow, however, if the transmitter and receiver plates are separate, they may need to be identical, particularly for narrowband applications.

The impact transmitter produces sound predominantly at the frequencies of the modes of vibration of the plate that are excited by the impact. The 'natural' resonant-mode frequencies of a plate are functions of the material, thickness, size, and shape of the plate, but also of how the plate is mounted in the transmitter body. The normal mode frequencies depend on the boundary conditions of the plate, such as at what edges the plate is mounted and whether the rim is rigidly edge-clamped or just fixed in place. Where the impactor strikes the plate will determine the linear combination of normal modes that will be excited.

Particularly in TWS applications, in which there is a premium on narrowband pulses to enhance S/N, it is important that the receiver plate be identical to the transmitter plate and that their mountings also be identical. The breadboard TWS impact-transducer system, shown in Fig. 1.4 and in a different view in Fig. 3.1, has a bandwidth of 7 Hz at a resonant frequency of 2186 Hz, for a fractional bandwidth of 0.003. With a narrow bandpass filter matched to the bandwidth (which we did not have), a mismatch of resonant frequencies of the transmitter and receiver plates of just 0.3% could cause a substantial degradation of the return signal that could mean the difference between detecting or overlooking a person.

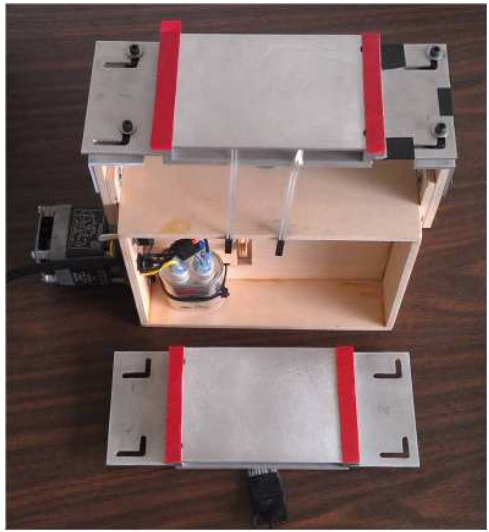

**Figure 3.1. Breadboard transmitter assembly and matched resonant receiver plate described in Sec. 8.**

In sonar applications, on the other hand, a wider receiver bandwidth may be preferred to allow for high-power operation or for Doppler shifting of signals received from targets. In some such cases, it may be desirable to have a narrowband transmitted signal, but a wider received bandwidth centered at the transmitter frequency. In other applications, such as wideband jamming, a wideband transmitter signal may be desired.

The preferred sensors for the receivers, whether in-air TWS sensors or underwater, are piezoelectric film sensors with attached leads and adhesive backing like transparent tape. Figure 3.2 shows the type of piezo-film sensors used in the breadboard TWS system described in Sec. 8, an FDT1-028K w/adh-F dart sensor with flexible lead attachment by Measurement Specialties. In Fig. 3.2, this piezo-film sensor is affixed to a 10-cm-dia, 1-mm-thick Al plate at a location where it is most responsive to the flexing of the fundamental mode. A good empirical way to optimize the positioning of the piezo-film sensors is by repeatedly measuring the in-band response of a sensor to a calibrated signal as the sensor is repeatedly lifted and re-taped onto new positions.

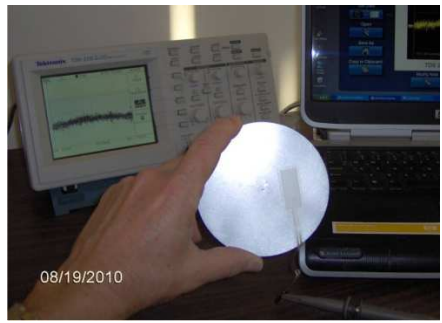

**Figure 3.2. Piezo-film sensors are affixed to the surface of the plate where the stress over the sensor surface is greatest for the vibrational mode to be measured.**

In general, a multiplicity of piezo-film sensors can be and should be attached to the (inside) surface of the thin plate of the receiver. The sensors should be attached with their polarities, corresponding to the vibrational mode that is being measured, in series. If *n* sensors attached in series each measure the same voltage response with the same polarity, the signal strength will be amplified by $n^2$, because signal strength scales as voltage squared. For example, the receiver plate in Fig. 3.1 has four piezo-film sensors mounted on its underside in a square array around the center of the plate. The flexible leads were soldered to each other in series at a 10-pin header (seen in Fig. 1.4), enhancing the signal voltage by a factor of 4 over a single sensor and the signal strength by a factor of 16.

For some vibrational modes, such as dipole modes, one piezo-film sensor may be stretched over a convex bend while another is compressed over a concave bend. In that case, their voltage polarities will be opposite each other, and their leads should be connected positive to positive or negative to negative in order for their voltage signals to add constructively.

If the receiver plate is not the same plate used for transmitting pulses, it should be identical, and the mounting of the receiver plate should be identical to the mounting of the transmitter plate. Even the bonding and fastening of the receiver plate to its mountings must be identical to the bonding and fastening of the transmitter plate to its mountings. Keeping the plates and the mountings identical helps to ensure that the normal mode resonant frequencies of both plates will be identical. The plates and their mountings need to be identical to a degree of accuracy such that the bandwidths of the useful resonances at the receiver substantially overlap the corresponding bandwidths at the transmitter.

Since the receiver plate differs from the transmitter plate in that it does not need to be attached to a more massive structure like the transmitter, as shown in Fig. 3.1, the transmitter plate should be acoustically isolated from the transmitter structure, so that the resonances of the transmitter structure do not couple to and affect the resonances of the transmitter plate. This isolation was accomplished, as described in Sec. 8, by rubber acoustic isolators.

## 4. Efficient Coupling of Transducers to Solid and Liquid Media

The central feature of the mechanical impact transducer is the use of a mechanical means, rather than an electromagnetic means, to produce high-power acoustic pulses in dense media, like water and walls, in the most energy-efficient way possible. To produce sound waves in any dense medium, whether water or a thick concrete wall, requires producing a physical motion of the medium. That is, a layer of the medium at its interface, much thinner than a wavelength, must be made to physically oscillate back and forth. The mechanical/impact transducer uses a mechanically induced oscillation of a source to produce a physical oscillation of the medium at the interface. A very efficient way to produce a high-power mechanical oscillation of the source with a low-voltage, low-power input is by a mechanical means, and one in which the source is well coupled to the medium.

This section discusses preferred methods of coupling mechanical impact transducers to walls and water. The method used to couple a transmitter to a wall can make a tremendous difference in performance, increasing the signal strength by orders of magnitude. With good coupling, the transmitter turns the wall to which it is coupled into a soundboard, like the soundboard of a piano, that vibrates at the same frequencies as the thin plate, but produces a louder sound because it moves a much greater volume of air. A good coupling of a transmitter to a wall amplifies the sound the same way holding a tuning fork to a tabletop amplifies the sound.

In a string instrument, like a piano or guitar, the strings vibrate against the soundboard usually via some sort of bridge. An effective 'bridge' to couple both the thin plates of the transmitter and the receiver to the wall was found to be two strips of 1-mm-thick, heavy-duty mounting tape fastened to two ends of the thin plate, seen as red strips in Fig. 3.1. The tape used here was Scotch® 411 Outdoor Mounting Tape. The 1-in-wide tape was cut lengthwise into 0.5-in-wide strips and applied to the ends of both the transmitter and receiver plates. The mounting tape was applied over the edge-clamped ends of the transmitter and receiver plates, and not over the active oscillating area, so that the tape would not affect the resonances when pressed or stuck to a wall.

During operation of the transmitter, the red plastic protective coverings of the two-sided tape did not need to be removed from the transmitter plate, but coupling of the transmitter plate to the wall seemed to be improved by applying 5 to 10 pounds of force to press the tape on the transmitter face against the wall. A conjecture that might explain this improvement is discussed below.

The bridging of the acoustic return signal from the medium into the receiver is as important as the bridging of the transmitted signal into the medium. An effective 'bridge' for coupling the sound waves from a wall into the thin plate of the receiver is similar to the 'bridge' for the transmitter. It is two strips of 1-mm-thick heavy-duty mounting tape at the two ends of the receiver plate, as seen in Fig. 3.1. Applying a force pressing the receiver against the wall did not seem to improve coupling appreciably, as it did for the transmitter, but receiver performance did seem to be improved by removing the tape coverings and adhering the receiver plate firmly to the wall. This particular tape is not reusable. In applications of an actual TWS device, methods to create reusable 'bridges' can be easily devised.

We have not yet systematically investigated to find the optimal coupling methods or optimal 'bridge' gaps. In each of our three acoustic TWS breadboards, represented in Figs. 1.2, 1.3, and 3.1, we settled for whatever coupling method worked sufficiently well to get results. Based on our experiences and the following analysis, however, we conjecture that the coupling efficiency of impact transmitters and receivers to walls may be optimized, respectively, by:
  (i)  Making the transmitter/wall gap as thin as possible without allowing the vibrating plate to contact the wall; and
  (ii) using a receiver 'bridge' that effectively transmits wall vibrations to the receiver unit without interfering with the free vibrations of the receiver plate.

Consider the transmission of a normally-incident plane sound wave from a solid plate into a solid wall across an air gap of width $L$, as illustrated in Fig. 4.1. The intensity transmission coefficient from the plate into the wall, conservatively neglecting forward reflections from the back surface of the plate, is [28]

$$T_I = 4\left[2+\left(\frac{Z_W}{Z_P}+\frac{Z_P}{Z_W}\right)\cos^2 kL + \left(\frac{Z_A^2}{Z_P Z_W}+\frac{Z_P Z_W}{Z_A^2}\right)\sin^2 kL\right]^{-1}, \quad (4.1)$$

where $Z_P$, $Z_A$, and $Z_W$ are the characteristic acoustic impedances of the plate, air gap, and wall, respectively, and $k$ is the wavenumber of the sound in air. Regardless of the plate and wall materials, because $Z_A \ll Z_P$ and $Z_A \ll Z_W$, the intensity transmission coefficient is

$$T_I = (2/kL)^2 Z_A^2 / Z_P Z_W, \quad (4.2)$$

for air gaps much narrower than a wavelength ($kL \ll 1$).

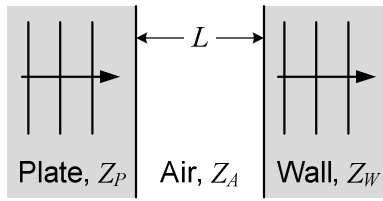

Plate, $Z_P$ | Air, $Z_A$ | Wall, $Z_W$

**Figure 4.1. Model for transmission of sound from transmitter plate into wall across air gap.**

Although the approximations render Eq. (4.2) unreliable, the scaling of transmission with air-gap thickness, $T_t \sim L^{-2}$, is to be expected. The difference between an air gap of 1 mm and 0.2 mm is 14 dB in transmission into a wall. The reason the transmitter worked better with a force of 5 to 10 pounds pressing the tape on the transmitter face against the wall is probably that the force squeezed the foam tape and the air gap to a thickness much less than the original 1-mm thickness. The air gap $L$ should not be made so thin, though, that the maximum amplitude of the plate vibrations $z_0$ causes the plate to contact the wall. That is, transmission into the wall is maximized for the narrowest gap subject to $L > z_0$.

For the TWS breadboard system illustrated in Fig. 1.2, we found the best coupling of the ceramic ultrasound transducers to the wall to be a layer of grease, like Vaseline or vacuum grease, between the transducer face and the wall. Pressing the bare transducers against bare walls without grease degraded performance. Since a force was generally applied to the COTS transducers to keep them in place against the wall, it may be that the grease provided a gap of order 100 microns that enhanced performance at those ultrasound wavelengths.

For water or other liquid media, an effective method of coupling impact transducers to is to have the vibrating plate drive the medium directly, as in the point design of Sec. 2. For an impact transmitter in a liquid medium, after the impactor has delivered a momentum impulse and kinetic energy to the transmitter plate, the primary avenue for that kinetic energy to be spent is by doing acoustic work on the medium.

When operated under water, the mode frequencies of a thin plate are downshifted, as was calculated in Sec. 2. The effect of the water is to increase the effective mass of the plate. For a given spring constant, a higher mass oscillates at a lower frequency. If the spring constant of the plate is stiff, then the water can be inconsequential. For example, the oscillations of a heavy-duty spring from the strut on a car would be unaffected by water. But if the spring constant of the plate is weak, then the water can significantly downshift the resonant mode frequencies.

From Eqs. (2.5) and (2.6), the effective mass $M_e$ and the spring constant of a thin plate, $K(\rho) = \omega^2(\rho) M_e(\rho)$, are nonlinear functions of the density of the medium in which the plate is submerged. In cases involving high radiation reactance, such as high density of the medium or low frequency of operation, a significant frequency downshift should be expected. System designs for underwater impact transducers should anticipate frequency downshifting, as in Sec. 2. Fortunately, as long as the transmitter and receiver plates are either the same plate or identical plates, the resonances of both plates will be frequency downshifted by the same amount, to first order.

## 5. Through-Wall Surveillance (TWS)

This section presents some design guidelines and operating parameters for acoustic transducers used in TWS to detect and track persons. The two primary advantages of acoustic TWS over radar are: (i) detection through metal and metal-lined walls; and (ii) detection of stationary persons through walls. Besides these, other attractive features of the acoustic TWS sensor and its signal processing include: high-resolution locating and tracking; portability; low cost; quick and easy preparation and deployment; near-real-time data processing and display; and no damage or changes to the wall. These features provide a robust stand-alone TWS capability or an excellent complement to a radar TWS sensor, as in the hybrid system explored in [19].

A TWS sensor that cannot detect stationary persons may be worse in some circumstances than no TWS sensor at all. A TWS sensor that can detect only moving persons can easily give the user a false confidence that he has accounted for all persons on the other side of a wall, which could make a life-or-death difference in certain military operations in urban terrain, law-enforcement, or first-responder situations. Particularly in scenarios involving clearing facilities, or in which dismounted infantry or law-enforcement personnel must enter facilities that may harbor militants or armed and dangerous persons, lives could be saved by the ability to detect people who are unconscious, sleeping, tightly bound, or otherwise immobile. Lives could even be saved of those trying to escape detection by remaining motionless, like stowaways inside cargo containers, who are at risk of death by dehydration on long voyages.

The acoustic TWS system uses discrete narrowband frequencies to detect phase changes in waves reflected from a moving person. By interfering successive return pulses, small changes in phase and amplitude within the reflected beam lead to big changes in voltage waveforms, allowing detection of mm-scale motion by cm-scale wavelengths. This very sensitive but simple means of detecting the mm-scale motion of persons who are breathing but otherwise stationary works *only* for narrowband frequencies.

The interference effects are washed out for wideband beams, like those of impulse radars, which are the most common radar TWS systems. These ultra-wideband (UWB) radars generally have a lower bound on detectable velocity of about 12 to 15 cm/s. For example, the Time Domain RadarVision 1000™ impulse radar could not detect velocities below 0.4 ft/s. Although the Raytheon MDRs used monotonic cw radar beams, they too were unable to detect velocities below 0.5 ft/s. According to developer Lawrence Frazier [9], "…we found that buildings breathed… In fact, a lot of things moved in the less than 1/2 a foot per second region… We filtered all that out. In fact, we basically operate in the 1/2 to 3 feet per second range."

Detecting stationary persons and locating and tracking moving persons through a wall requires specialized software algorithms to process the sound waves scattered from everything on the other side of the wall and received at multiple receivers. In 2000, we developed a locating and display algorithm that converts the voltage waveforms from two receiver channels into a map pinpointing the locations of multiple persons behind a wall. The algorithm, presented in Sec. 6, is fast, accurate, robust, and easy to implement and use. Figure 1.2 showed an actual implementation of this algorithm on real TWS binaural waveforms.

Depending on the angular divergence of the transmitter beam, an array of transmitters might be needed to provide TWS coverage over a large area. The divergence of the transmitted beam may not be much larger than the full-angle diffraction-limited divergence of $1.22\lambda/a$, where $\lambda$ is the wavelength in air and $a$ is the effective radius of the transmitter plate. As explained in Sec. 2, $a$ can be substantially smaller than the plate radius $b$, and the divergence substantially wider, because only the central portion of the plate contributes significantly to beam power.

We have developed concepts for linear arrays of transmitters and receivers in through-wall sensors. For wide-area coverage, probability of detection is generally greater with multiple transmitters than could be achieved with a single transmitter scaled up in power. Because the cost and weight and size of mechanical transmitters are all low, using an array of mechanical transmitters is worth considering. Depending on divergence, a linear array of transmitters spaced a few meters apart laterally and 75 to 100 cm above the floor should provide thorough room coverage and should maximize the probability of detecting persons lying down, sitting, and standing.

An array of transducers would either need to be tagged or sequentially phased in time for the receiver processors to be able to identify the source of the reflected signal for triangulation. The transmitters could be tagged, or identified, by different frequencies, but such a method would be costly and processing-intensive. Alternatively, identical transmitters could be triggered sequentially in a predetermined pattern. Regardless of which receiver detects an echo signal, that signal would be referenced to the transmitter that produced it. A 10-m maximum range requires a minimum delay between pulses of different transmitters of about 60 ms, so there is no overlap of pulses at receivers. The maximum allowable pulse repetition frequency of the non-overlapping sequential pulses, about 16 Hz, is more than sufficient for TWS.

Owing to the processing algorithms used, jitter in the start of each pulse of a mechanical transmitter is inconsequential. What we have found to be critical to processing, instead, is that the start time of each pulse be known with an uncertainty, $\Delta t$, less than about 1/20 of a wave period, or ideally one sample time, at most. Section 7 presents a means of achieving such a low $\Delta t$ with mechanical transmitters that was used in demonstrating the breadboard TWS system in Fig. 3.1. In simple terms, the concept for achieving low $\Delta t$ is the following.

In Fig. 3.1, one sees leads from two piezo-film sensors that are affixed to the underside of the transmitter plate. When the impactor strikes the transmitter plate, the piezo-film sensors instantaneously produce a sharp voltage spike of tens of volts that establishes the fiducial $t = 0$ with extremely low uncertainty. (One can choose the location of a piezo-film sensor on the transmitter plate to produce a fiducial voltage spike of any lesser voltage, as desired.) In our breadboard demonstration, the voltage spike, unamplified of course, was used to trigger the receiver. The shot-to-shot reproducibility of the start time determination by this method was as accurate as could be measured, that is, within one sample time.

## 6. Locating and Tracking Display Algorithms

This section discusses the 3-step algorithm used to produce a TWS *x-y* display that shows the locations and that tracks in near-real-time all persons on the other side of a wall. When implemented properly, the algorithm transforms binaural voltage waveforms, like those seen on the left side of the screen in Fig. 1.2(b), to the map of the current locations of persons, like that seen on the right side. (We note again that the lack of cross-range resolution in this figure is not caused by the algorithm, but is caused by inadequate lateral spacing between receivers, itself a consequence of the limited capabilities of COTS transducers.)

The 3-step algorithm is so fast because it emulates the way the human mind and eye perceive complex waveforms of thousands of data points. For example, looking at the left side of the screen in Fig. 1.2(b), which represents the voltage waveforms *vs*. time in both receivers, one quickly notices that there are two periods of time in both waveforms when the waveform envelopes seem to exceed the general noise level. If the S/N of the COTS transducer system were not so poor, the exceedances would be more pronounced. These two periods of exceedances in each waveform represent reflections from the two persons, who are sitting still. Just as the eye quickly picks out the periods of exceedances in the voltage waveforms, the algorithm is designed to do the same.

The 3-steps of the algorithm are:
(i)   <u>Filter</u> noise
(ii)  <u>Count</u> data points in time bins
(iii) <u>Triangulate</u>

To demonstrate this algorithm, we have simulated stereo-receiver voltage waveforms <u>after the dc component has been subtracted out</u> and with assumed automatic gain control, which increases the gain on a target corresponding to its range. The simulated stereo voltage channels are shown in Fig. 6.1. Time on the *x*-axis is in ms and measurements are in feet.

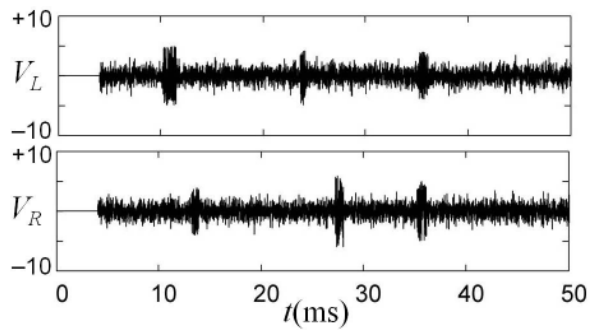

**Figure 6.1. Simulated left and right channels of stereo ultrasound TWS waveforms (volts) *vs*. time (ms).**

The waveforms represent typical ultrasound return signals with a 50-ms record length beginning 4 ms after receiver triggering and a 100-kHz sampling rate. There are 5000 data points in each channel. With this record length and a 4-ms gate, the range of each receiver is 2.3 ft to 30.5 ft. With this range, stereo receivers separated by 5 ft at the center of one wall are capable of triangulating anywhere in a 25-ft x 25-ft room (except for the immediate proximity of the receivers).

The first step of the algorithm is to filter out noise. Any data point with a value less than an adjustable voltage threshold $V_0$ is set equal to zero. On a device in use, the adjustable threshold $V_0$ should be tunable with a 'brightness' dial.

The second step of the algorithm is to count how many data points remain in each time bin after filtering. The record length is divided into a number of equal time bins for this purpose. If a resolution of 0.5 ft x 0.5 ft is desired in the *x-y* display, then we have found that time bins of 0.5 ms provide adequate resolution, but time bins of 1 ms do not. If the time bins are too big, then the *x-y* display may overlook people and show ghost images.

Figure 6.2 shows the histograms produced from the waveforms of Fig. 6.1 by performing Steps (i) and (ii) of the algorithm. The data in Fig. 6.1 were filtered with a brightness threshold of $V_0$ = 3.2. Then the data points remaining after filtering in each of 100 0.5-ms bins were counted. Figure 6.2 shows the results of the counting.

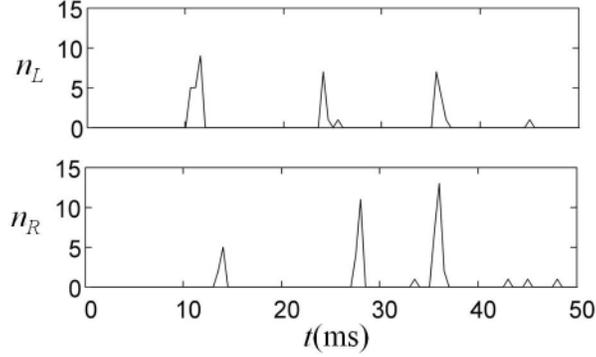

**Figure 6.2. Numbers of data points of each stereo channel remaining in each 0.5-ms time bin *vs*. time (ms) after filtering of waveforms in Fig. 6.1 (at $V_0$ = 3.2 volts).**

The third step of the algorithm is triangulation, which involves transforming from range coordinates to *x-y* coordinates and correlating the filtered data in each *x-y* cell. The third step is the most processing-intensive, but it involves working with time bins, rather than data points, and there are 10,000 times fewer time bins than data points in this example.

To implement the third step, an *x-y* grid is established, with $p$ labeling the cells in the *x* direction, and $q$ labeling the cells in the *y* direction. The value assigned to each cell $(p,q)$ is the product of the count from the 'appropriate' time bin of the left receiver with the count from the 'appropriate' time bin of the right receiver.

Suppose $P0_m$ is the count of data points from Step (ii) in the $m^{th}$ time bin of the left receiver, and $P1_n$ is the count in the $n^{th}$ time bin of the right receiver. Suppose that the time bins are 0.5 ms, the *x-y* cells are 0.5 ft x 0.5 ft, the left receiver is 10 ft (20 cells) from the left corner of the room (where $p = 0$), and the right receiver is 15 ft (30 cells) from the left corner of the room. Then the correlated-count value assigned to each cell $(p,q)$ is

$$C_{p,q} = (P0_m)(P1_n), \tag{6.1}$$

where the 'appropriate' time bins are

$$m = \text{round}\left\{K[(p-20)^2 + q^2]^{1/2}\right\}, \tag{6.2a}$$

$$n = \text{round}\left\{0.5K\left([(p-20)^2 + q^2]^{1/2} + [(p-20-2d)^2 + q^2]^{1/2}\right)\right\}. \tag{6.2b}$$

Here, "round" is a function that rounds off numbers to the nearest integer; $K \equiv 2/c = 1.77$ is a constant that accounts for the double round-trip time at the speed of sound, $c = 1.13$ ft/ms; and $d$ is the separation between receivers on the $x$ axis.

After this third step of the algorithm was performed on the data in Fig. 6.2, the correlated-count value, $C_{p,q}$, is shown as a 3D contour plot in Fig. 6.3. The grid numbers in Fig. 6.3(a) are coordinate distances in feet. The left and right receivers are located along the $x$ axis at $x = 10$ ft and $x = 15$ ft, respectively.

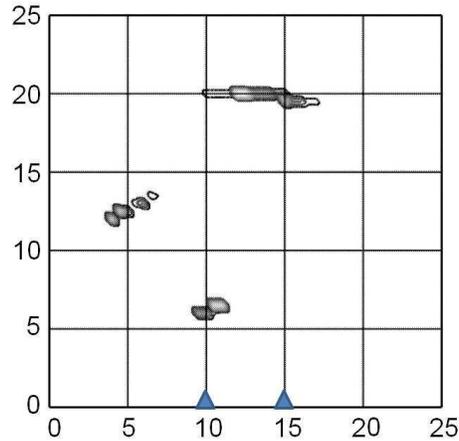

**Figure 6.3. Contour plot of the processed data in Fig. 6.1, showing the location of three persons in a 25-ft x 25-ft room. Receivers are on $x$ axis at 10 and 15 ft. Brightness is $V_0 = 3.2$.**

The 3-step algorithm is fast, because it emulates human perception in picking out only clusters of data points that stand out above the noise. Then, instead of correlating all of the data points in each channel (25 million correlations in this example), it performs only one correlation per $x$-$y$ grid cell (25 hundred correlations).

The accuracy of the algorithm in locating persons is limited by the pulse detection error, which is of the order of the return pulse width. The algorithm reduces the numerical resolution to this low level of accuracy (a 0.5-ms time bin in this example) before correlating the two channels. Thus, the algorithm does not need to degrade the realizable locating accuracy. One can see in Fig. 6.3 that the resolution is better than 1 ft at ranges comparable to the receiver separation, and the cross-range resolution is degraded only at ranges much longer than the separation.

The algorithm is robust at detecting signals even at low S/N, as was demonstrated in Fig. 1.2, and at eliminating false alarms (ghost images), as long as the time bins are chosen small enough (0.5 ms in this example). In Fig. 6.1, two of the targets had $S/N \approx 4$, yet were easily detected and located by the algorithm. The algorithm is even robust at discriminating closely spaced persons.

The algorithm will be easy to use in a TWS device. The only parameter to be varied by the user is brightness (filter threshold $V_0$), which may be controlled by a dial, with immediate visual feedback to the user.

## 7. Signal Processing for TWS

The purpose of this section is to explain how an acoustic TWS sensor:
  (i) can have S/N improved up to about 30 dB through a well-designed filter/amplifier; and
  (ii) can be made to detect and locate humans in near real time.

The voltage waveforms acquired from each of the receivers and their sensors by the methods outlined in Sections 3 and 5 are processed through the following steps, illustrated in Fig. 7.1:
  (i) receiver signal triggered by transmitter (after fixed delay);
  (ii) voltage waveforms digitized;
  (iii) $t = 0$ correlated to $t = 0$ of average of past $n$ waveforms;
  (iv) narrowband filter applied;
  (v) filtered waveform subtracted from average of past $n$ filtered waveforms;
  (vi) amplifier applied with automatic gain control (AGC);
  (vii) for optional tracking, 3-step locating and tracking display algorithm of Sec. 6 applied, including: (a) filter noise; (b) count data points in time bins; and (c) triangulate.

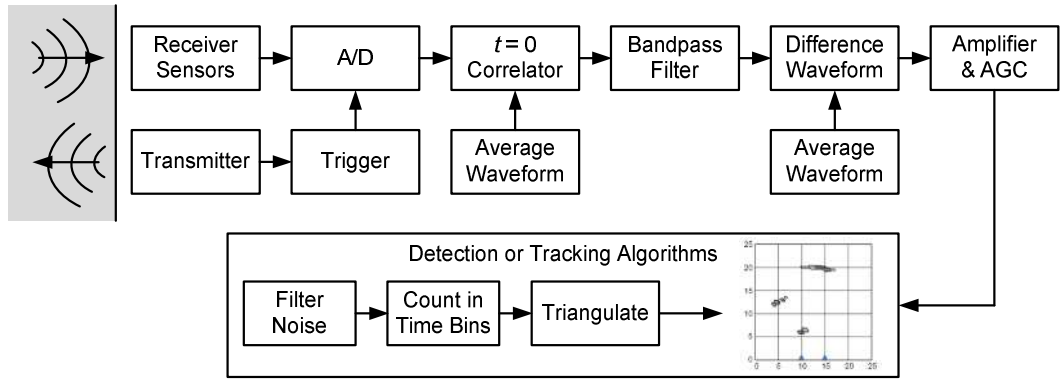

**Figure 7.1. Block diagram of steps for processing echoes at receivers into an *x-y* map of persons on the other side of a wall.**

These signal-processing steps are explained more fully below. Then a detailed example of these steps will be given for an actual detection made by the breadboard impact transducer described in Sec. 8.

The signal processing for each transmitted pulse begins with the impactor striking the transmitter plate. That event triggers by any means the start of a delay period of some number of milliseconds. For example, to trigger the delay period, the impactor can produce a large voltage spike in a piezo-film sensor affixed to the plate, as described in Sec. 5. The delay period is needed so that the sound pulse produced by the impact and conducted through the air and through the walls to the receivers will be sufficiently diminished at the receivers by the end of the delay period. The downside of the delay period is that persons in the immediate vicinity of the transmitter will not be detected. And the longer the delay period, the greater the 'dead zone' for detection.

After the delay period ends, the signal processor begins acquiring voltage waveforms from the sensors on the receivers. If it is only required to detect humans, and not to locate or track them, then only one receiver may be necessary. Otherwise, a horizontal linear array of at least two receivers is needed, as discussed in Sec. 6. The voltage waveforms from each receiver are then digitized.

The digitized waveforms carry information about the acoustic waves reflected from all objects and persons on the other side of the wall. (Since the receivers may also receive reflections from the same side of the wall, it may be necessary to operate the system remotely if transmitter back lobes and side lobes cannot be suppressed.) The next steps in signal processing are to remove reflections from completely motionless objects.

At all times, a running average is kept in memory of the past $n$ waveforms from each receiver. Since the noise level will be suppressed and S/N will be increased by a factor of about $n^{1/2}$, the number of waveforms in the average should be as large as allowed by operational constraints. Most likely, if the triggering technique of Sec. 5 is used, the newest waveform will start at exactly the same time delay with respect to the impactor striking the plate as the average waveform. Or at least the difference in time delays with the average waveform will be very much less than the wave period. If not, then it may be necessary to adjust the start of the newest waveform to correspond to the start of the average waveform. This can be done by cross-correlating the newest waveform with the average waveform.

After $t = 0$ has been established for the newest voltage waveform for each receiver, a narrow bandpass filter is applied to the waveforms. The filter should be matched to the resonant frequency or frequencies of the transmitter and receiver. (The resonances should be identical in both.) And the bandwidth of the filter should be matched to be no less than the full width at half maximum (FWHM) of the spectral resonances. As long as the bandwidth encompasses the FWHM, the narrower the bandpass filter, the more noise is excluded from the signal and the higher the S/N. S/N increases at the margin almost in inverse proportion to the bandwidth, but the bandpass filter should not be so narrow that it excludes a significant part of the in-band resonant energy, or that it reduces probability of detection, as discussed later in this section.

After the waveforms have been filtered, the filtered waveform from each receiver is compared to, and subtracted from, a running average of the past $n$ filtered waveforms from that receiver. That is, the newest filtered waveform is destructively interfered with the most recent average filtered waveform.

The output of that waveform interference process, the difference waveform, should only differ from noise for those reflections from targets that are not motionless. The difference waveform is amplified with an automatic gain control (AGC). The AGC is designed to compensate for the weaker reflections arriving from more distant targets. The power of the diffuse sound waves reflected from a human target scales with range $r$ as about $r^{-3.8}$ [5]. The time-of-flight of the reflected sound waves is measured, and the range is half the time-of-flight times the sound speed.

The filtered and amplified difference waveform from one receiver may be sufficient to <u>detect</u> persons through a wall. To <u>locate and track</u> persons requires at least two receivers. The algorithms for locating and tracking multiple persons through a wall with two receivers were presented in Sec. 6. If the transmit and receive beam divergence is narrow relative to the area to be scanned, then a linear array of more than two receivers and/or transducers may be needed, as discussed in Sec. 6. In that case, difference waveforms can be created from neighboring pairs of receivers, or with some changes to the tracking algorithms, difference waveforms can be created from more than two waveforms, with some improvement in resolution, particularly cross-range resolution.

As an example of the application of these signal processing techniques, we consider actual waveforms produced and measured by the breadboard impact transducer and matched receiver system shown in Fig. 3.1. The transmitter and receiver plates were identical, and their resonance at $f_0 = 2186$ Hz was used. The quality factor was $Q = 343$, suggesting that the full width at half-maximum (FWHM) intensity of the presumed line shape was about 7 Hz, for a fractional bandwidth of 0.003. The power spectral density (PSD) of the transmitter or receiver plate is shown in Fig. 7.2. A COTS bandpass filter of 1 to 10 kHz was applied to the PSD in Fig. 7.2.

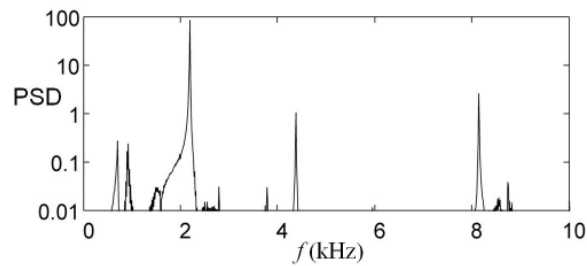

**Figure 7.2. PSD (arb. units) vs. frequency (kHz) for strike test on receiver plate assembly.**

The signal processing for the breadboard TWS system did not have a narrowband filter/amplifier or the capability to produce difference waveforms in real time. Virtually all the results that follow, therefore, were derived in post-processing. Without a narrowband filter/amplifier that would have increased S/N by about 26 dB, as discussed below, the breadboard system could only detect humans between two 'walls' along a one-way path from the transmitter to the receiver. The two 'walls' were a massive mahogany desk, which acted as a soundboard (see Sec. 4), and a solid hardwood table, nearly 1-in thick and over 5 meters away.

Figure 7.3 shows a comparison in post-processing of three successive through-wall tests, in which the received waveforms were each averaged over 16 pulses by a recording oscilloscope. First, a baseline voltage waveform (fine black in Figs. 7.3(a) and 7.3(b)) was created with the operator motionless. Then with the operator position unchanged, a second waveform was created and overlaid in post-processing as the light-gray waveform in Fig. 7.3(a), covering most of the baseline waveform. Lastly, a third waveform was created as the operator moved about between the two 'walls', and was overlaid as the gray waveform in Fig. 7.3(b). The heavy black waveforms in both Figs. 7.3(a) and 7.3(b) are the differences between the gray and fine black waveforms.

With the transmitter and receiver about 18 feet apart, after the transmitter triggered the receiver, there was a delay of over 15 ms before the acoustic waves from the transmitter accumulated at the receiver to produce a signal distinguishable from noise, as seen in Figs. 7.3. In Fig. 7.3(a), the difference waveform (heavy black) is almost the same as the noise level before the acoustic waves are noticeable at the receiver. In Fig. 7.3(b), the effects of the changed position of the person, evident as a much larger difference waveform (heavy black), are seen throughout the pulse owing to multipath propagation of sound in a cluttered environment.

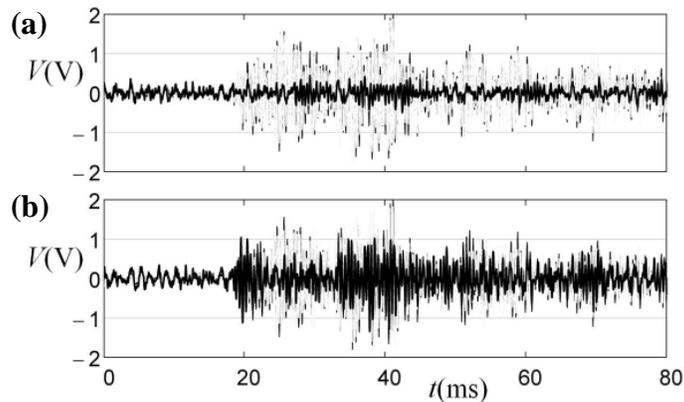

**Figure 7.3.** Voltage (V) vs. time (ms) for baseline waveform (fine black) and difference waveform (heavy black) with: (a) second waveform (gray) with operator in same position; and (b) third waveform (gray) in changed position.

The PSDs of the received signals cannot distinguish between moving and stationary persons as well as the difference waveforms can, because the frequency content of the received signals is not changed nearly as much as their phase. The differences in the frequency content do become significant, however, when comparing the inverse fast Fourier transforms (IFFTs) in a filtered band about the resonance.

Figures 7.4 and 7.5 compare the IFFTs in filtered bands about the resonant frequency. In these figures, numerical bandpass filters were emulated by only performing the IFFT on the Fourier frequency components within a specified frequency interval. In Fig. 7.4, the IFFT was performed only on the three neighboring Fourier components at 2186, 2198, and 2211 Hz. These three components span a filter bandwidth that is 1.1% of the central frequency of 2198 Hz. In Fig. 7.5, the IFFT was performed on the 11 Fourier components from 2101 Hz to 2223 Hz, which span a filter bandwidth that is 5.6% of the central frequency of 2162 Hz.

The purpose of a narrowband filter/amplifier is to exclude all noise outside the narrow bandpass filter, but still admit most of the useful signal, and thereby increase the S/N. The COTS filter/amplifier used in this breadboard demonstration had a bandpass filter from 1 kHz to 10 kHz. That means it admitted all noise over a 9 kHz range. But, as seen in Figs. 7.4 and 7.5, most of the useful signal is contained only within about a 24 Hz to 120 Hz range. In this example a properly designed filter/amplifier, therefore, could eliminate of the order of about 19 dB to 26 dB of noise from the filtered signal, improving the S/N by like amounts.

This estimate of S/N improvement from a narrowband filter/amplifier is confirmed by the PSD of the noise spectrum. The noise spectrum was filtered and amplified through the same filter/amplifier that was used for the measurements in Fig. 7.3. That is, the bandpass filter was set from 1 kHz to 10 kHz and the gain was set to 40 dB. The noise power integrated over the entire spectrum was 8.53 (in arb. units). The noise power within the 5.6% band of Fig. 7.7 was 0.134 in the same units. The gain in S/N to be achieved by a 5.6% bandpass filter by these estimates, therefore, is about 18 dB, and about 25 dB by a 1.1% bandpass filter.

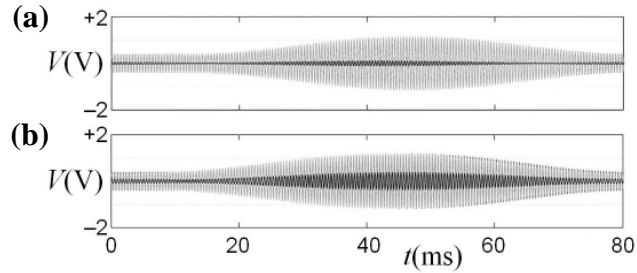

**Figure 7.4. IFFT voltage (V) with 1.1% bandpass filter vs. time (ms) for baseline waveform (fine black) and difference waveform (heavy black) with: (a) stationary-person waveform (gray); and (b) moving-person waveform ( gray).**

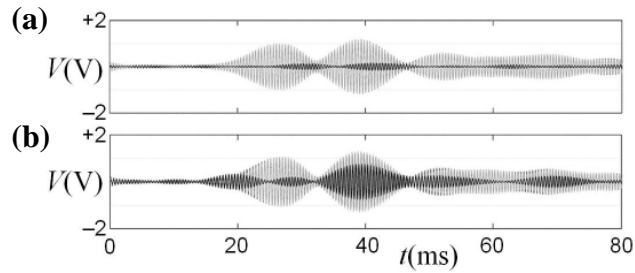

**Figure 7.5. IFFT voltage (V) with 5.6% bandpass filter vs. time (ms) for baseline waveform (fine black) and difference waveform (heavy black) with: (a) stationary-person waveform (gray); and (b) moving-person waveform ( gray).**

From Figs. 7.4 and 7.5, one sees that if the bandpass filter is too narrow, the distinction between a moving person and a stationary person fades. The distinction between a moving person and a stationary person with a 1.1% bandpass filter in Fig. 7.4 cannot be made with as much confidence as the distinction with a 5.6% bandpass filter in Fig. 7.5. The optimal bandwidth of the bandpass filter, therefore, is a tradeoff between elimination of noise with narrow filters and confidence of detection with wider filters. The bandpass filter should be wide enough to encompass more than a few Fourier components of the FFT. In practical terms, that condition is equivalent to the condition that the bandpass filter should be much wider than the inverse of the record length.

For operational reasons, a prf of a TWS sensor may need to be at least about 3 Hz, which would limit the record length to a few hundred ms. Then the bandpass filter should be much wider than a few Hz, say, at least a few tens of Hz. For the breadboard TWS system, the record length in Figs. 7.3, 7.4, and 7.5 was 81.92 ms, suggesting that a bandpass filter for the breadboard should be at least about 100 Hz wide, or about 5% of the operating frequency of 2.2 kHz. This condition was confirmed by the comparison between Figs. 7.4 and 7.5, which showed a loss of detection capability with a 1.1% bandpass filter. As seen in Figs. 7.4 and 7.5, the frequency sideband at 2125 Hz, which would have been missed by a 1.1% bandpass filter, is significant in distinguishing moving and stationary persons.

Filters that are too narrow should also be avoided because the resonant frequencies can change or drift, depending on many conditions. The resonant frequency of the breadboard receiver, when it is held in the hand and struck like a tuning fork, is 2186 Hz. The resonant frequency when it is fastened by mounting tape onto a wall, as in Figs. 7.4 and 7.5, is 2198 Hz, a difference of 0.5%.

To determine the optimal bandwidth and central frequency of a bandpass filter, the system should be tested under a wide range of operating conditions, varying the:
  (i)   wall types;
  (ii)  separation of transmitter and receiver;
  (iii) mounting of the transmitter and of the receiver;
  (iv)  transmitter/wall air gap;
  (v)   impact power of the impactor on the plate (separation of plate from impactor);
  (vi)  range to target (higher frequencies attenuate faster);
  (vii) and any other factors that might affect the spectral response of the receiver.

As for the amplifier part of the custom filter/amplifier, additional amplification will be needed for detection of humans through heavy-duty steel walls and over distances of as much as about 10 m. Ideally, the gain should automatically increase as a function of time after plate impact, as discussed above.

Figure 7.6 shows the improvement to S/N and signal processing that can be gained by a properly designed filter/amplifier combination. The MCSC transmitter shown in Fig. 1.3 was tunable within a range of about 8 to 10 kHz. A custom 8 – 10 kHz bandpass filter/amplifier was built and used with this transmitter. Figure 7.6 shows a raw measured PSD (in red) of the MCSC transmitter through a wall, and the same PSD after it was filtered and amplified (in blue). The in-band signal gain is 60 dB, and the noise floor is increased by 30 dB, for a net gain of S/N of 30 dB.

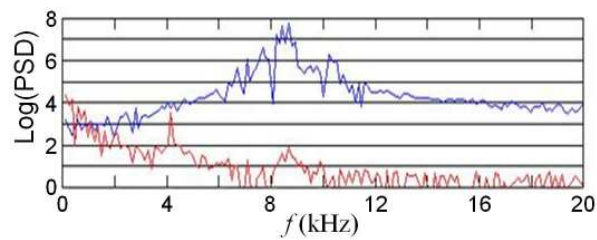

**Figure 7.6.  Log PSD (arb. units) vs. frequency (kHz) of MCSC transmitter with 8–10 kHz bandpass filter and 60-dB gain (upper) and without filtering or gain (lower).**

The breadboard TWS system demonstration described in the next section had only rudimentary signal processing capabilities and required post-processing for human detection. The following discusses how the breadboard signal processing would need to be augmented for actual TWS applications, like scanning cargo containers for detection of stowaways, to provide near-real-time signal processing capability and an increase of S/N by several tens of dB.

Because the operating frequency of the breadboard transducer system, less than about 2.2 kHz, is relatively low, the sampling rate requirements are modest. In order to accommodate round-trip travel times of acoustic waves in cargo containers, most of which are 40-ft (12-m) long, especial-ly when the containers might be cluttered and the acoustic waves might take some time to reverberate and accumulate at the receiver, the record length should be at least about 80 ms and need not be more than about 160 ms. According to the Nyquist criterion, the lowest sampling rate that could possibly measure the resonance is about 5 kS/s, but such a low rate would lead to aliasing errors. Better is a sampling rate of 10 kS/s, and more than adequate is a sampling rate of 25 kS/s.

The signal processor for a TWS product should therefore be designed to accept the signals from a narrowband filter/amplifier at a pulse repetition frequency (prf) of at least 3 Hz. (This prf is about the quickest that allows the reverberations from the previous pulse to decay.) Following are the signal-processing steps that will need to be performed with the waveform output from the bandpass filter:
  (i) The $t = 0$ mark on the waveform should be established by the methods described above.
  (ii) The voltage waveform should be centered about the mean voltage.
  (iii) The voltage waveform should be stored in memory.
  (iv) A running average of the past $n$ waveforms, where $n$ is 10 to 16, should be constantly updated in memory. The new $(i+1)^{th}$ waveform should be subtracted from the running average of the past $n$ waveforms, up to and including the $i^{th}$ waveform, in order to produce a difference waveform.
  (v) If this difference waveform exceeds a threshold voltage over multiple contiguous samples, that indicates a detection.
  (vi) Every time the new $(i+1)^{th}$ waveform is added to the average, the $(i+1-n)^{th}$ waveform should be removed from the running average of the past $n$ waveforms to update the running average.
  (vii) At a prf of 3 Hz, the first real-time detection can be made within about 3 to 5 seconds.

This method of real-time detection by exceedances of a threshold voltage is illustrated in Fig. 7.7. The data from the baseline, second, and third waveforms shown in Fig. 7.3 were used with a thresholding algorithm that only displays voltage exceedances above 0.5 V. Since the noise amplitude is almost entirely below 0.5 V, voltage differences above this threshold caused by human

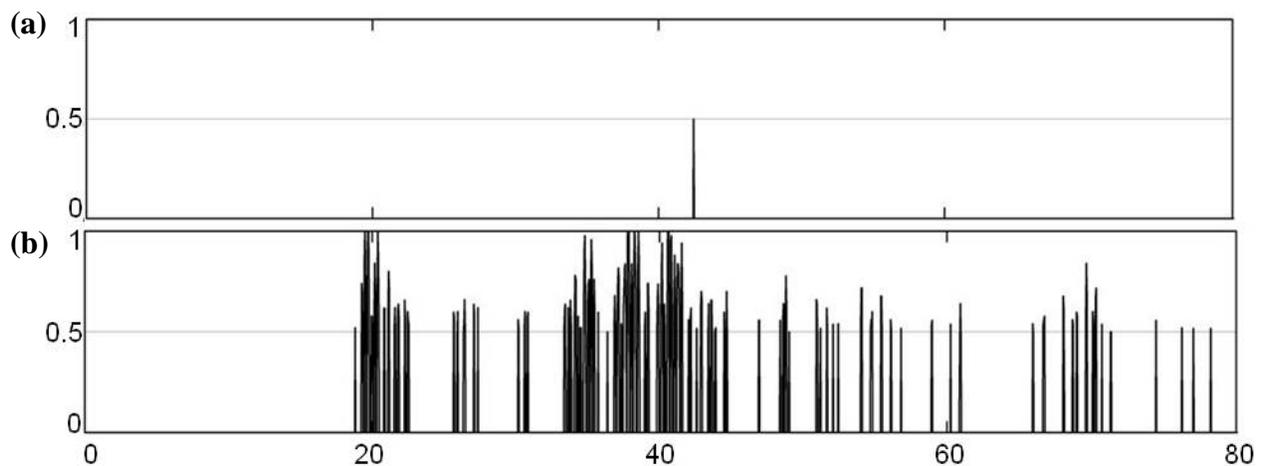

**Figure 7.7. Threshold-detection display (volts *vs.* time in ms), based on difference waveform data from Fig. 7.3, for: (a) stationary person; (b) moving person.**

motion are readily apparent. (The tapering-off of exceedances at late times would not occur with an AGC.)

Noteworthy about Fig. 7.7 is that the detection of a moving person was accomplished without a narrow bandpass filter or FFTs or IFFTs. The detection involved just a subtraction of waveforms and amplification of the difference waveform without even an AGC, suggesting that just rudimentary signal processing may be adequate for TWS with a sufficiently powerful transducer, such as the impact transducer used here. If tracking is needed, however, and not just detection, then some of the signal processing methods discussed here, and particularly AGC, may be necessary for the tracking algorithms to work well.

## 8. Breadboard Impact Transducer for TWS

This section describes the design and characterizes the performance of the breadboard TWS transducer system that was used to demonstrate the detection (with only rudimentary signal processing, in post-processing) of a person moving between two thick solid walls over 5 m apart, as seen in Figs. 7.3 and 7.7. Figure 8.1 shows the complete breadboard TWS transducer system.

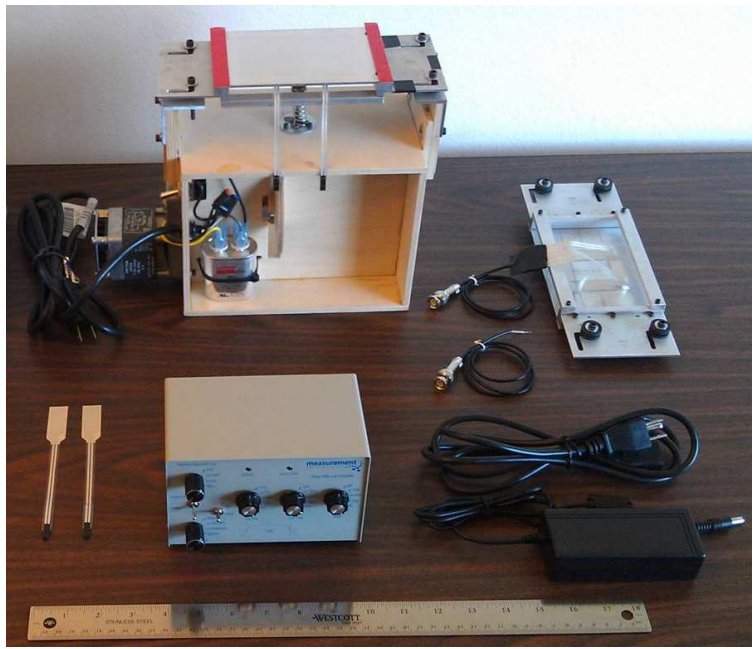

**Figure 8.1. Complete breadboard TWS transducer system, size relative to 18-in ruler.**

From left to right and back to front, the items in Fig. 8.1 are:
- Transmitter with attached:
    - Motor, capacitor starter, and power cord
    - Actuator
    - Transmitter plate assembly with attached:
        - 4 Mounting bolts and acoustic isolators
        - 4 Piezo-film sensors

- 2 Heavy-duty ½" strips of mounting tape
- Resonant receiver plate assembly with attached:
    - 4 Mounting bolts and acoustic isolators
    - 2 Heavy-duty ½" strips of mounting tape (not shown)
    - 4 Piezo-film sensors, connected in series to 10-pin header
    - BNC cable soldered to 10-pin header.
- BNC cable with bare-wire ends, for triggering by transmitter
- 2 Spare piezo-film sensors with flexible leads and self-adhesive backing
- COTS filter/amplifier with transformer and power cord

To repeat the simple demonstration of detection with post-processing that resulted in Figs. 7.3 and 7.7 using this equipment, the following steps would be performed:
1. Place the transmitter about 15 to 25 feet from the receiver and COTS filter/amplifier.
2. Place the transmitter, with the transmitter face down, on a solid table or other wall-like surface. Do not remove the red plastic covers from the foam mounting tape. Instead, place a weight of about 5 to 10 pounds on the transmitter, so the covered mounting tape is squeezed firmly against the table top.
3. Set the oscilloscope to be triggered directly by each transmitter pulse.
4. Test the receiver by tapping the center of the face of the receiver plate very lightly with a metal object. The voltage waveform should look like that in Fig. 8.2.
5. Remove the red plastic covers from the two strips of mounting tape on the face of the receiver plate, and place the receiver plate assembly face down on a table or a wall, so the mounting tape sticks to the surface, next to the filter/amplifier and oscilloscope. (Note: A little grease on the table or wall surface will make the tape easier to remove, but the tape should make good contact with the surface.)
6. Start the transmitter and, remaining still, record two waveforms after averaging for 16 pulses each. The waveforms, when overlaid, should look like Fig. 7.3(a).
7. Change position near the receiver and record a third waveform after averaging for 16 pulses. The waveform, when overlaid with either of the first two waveforms should look like Fig. 7.3(b).

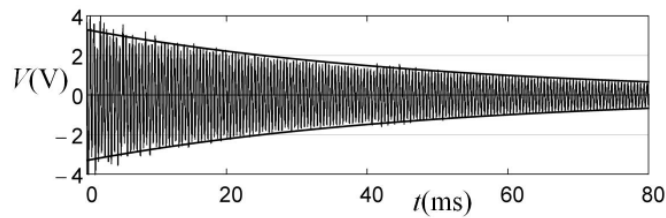

**Figure 8.2. Voltage (V) vs. time (ms) for strike test on receiver plate assembly. Curves are best fit of exponential decay to waveform envelope.**

In Fig. 8.2, the best fit of the envelope of the amplitude of the voltage oscillations having the form $V = A\exp(-at)$ is for $A = 3.3\,\text{V}$ and $a = 20\,\text{Hz}$. Since the quality factor $Q$ is given in terms of the decaying voltage waveform by $Q = \pi f_0 / a$, where $f_0 = 2186\,\text{Hz}$ is the resonant frequency of the receiver plate, we find $Q = 343$.

This estimate of $Q$ is supported by an estimate of the line width of the resonance. Figure 7.2

shows the PSD of the voltage waveform in Fig. 8.2. Since the sampling rate in these measurements was 25 kHz and the record length of the waveform for the FFT was 2048 samples, the frequency resolution of the PSD was only 12.2 Hz. Zooming in, as in Fig. 7.4(b), on the three frequencies about the resonance at 2186 Hz in Fig. 7.2, however, suggests that the FWHM intensity of the presumed line shape was about 7 Hz, for a fractional bandwidth of 0.003, consistent with the measurement of $1/Q = 0.0029$ above.

The breadboard impact transmitter works as follows. The actuator for the impactor is a spring pullback and release mechanism, shown in Figs. 8.3 and 8.4. The steel impactor is welded to its post and to the top of the spring via a cylindrical mounting inside the spring, as in Fig. 8.3. The post is pulled back against the spring by a cam follower that follows the spiral arc of the rotating cam seen in Fig. 8.3(a).

When the cam follower has compressed the spring to its maximum amplitude, the cam follower falls over the lip of the cutout in the cam, seen in close-up in Fig. 8.4(b). The sudden release of the cam follower by the cam releases the post to which the impactor is welded, and the compressed spring is 'unlatched' to drive the impactor against the underside of the transmitter plate, seen in Fig. 8.3(b). The cycle then repeats, with the outer edge of the rotating cam smoothly picking up the cam follower at small radius and pulling the cam follower to larger radius against the increasing force of the spring.

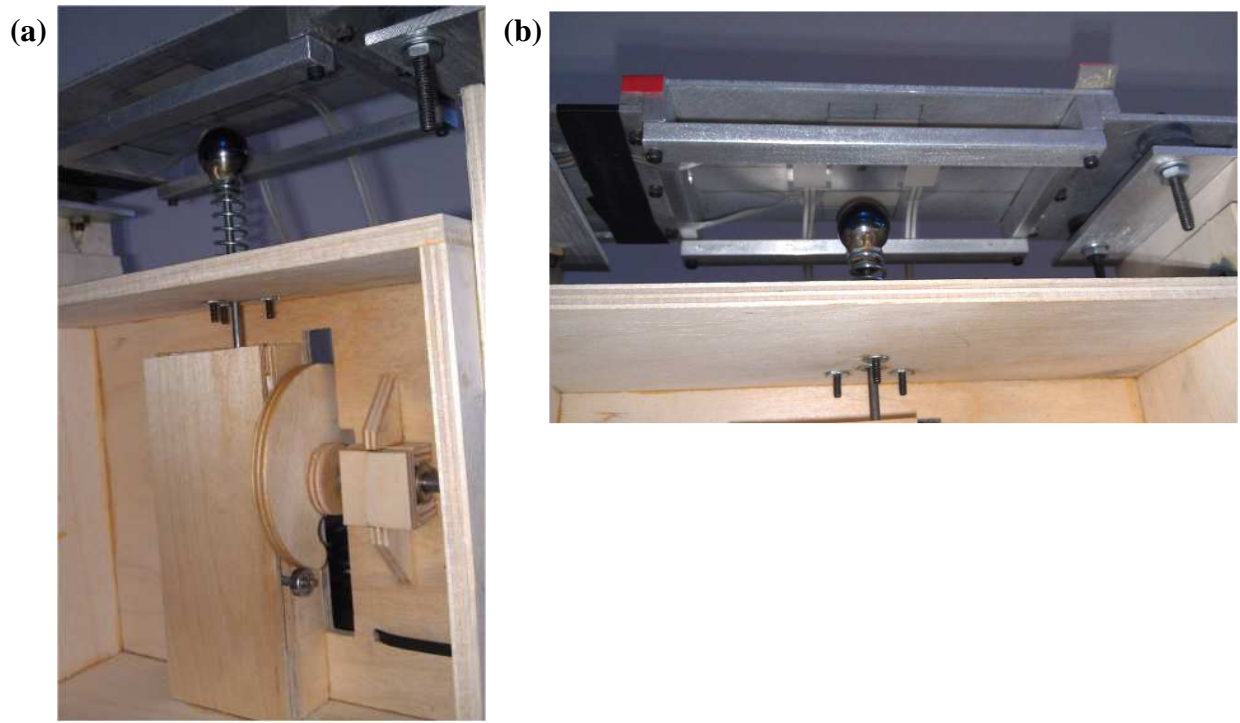

**Figure 8.3.** Transmitter: (a) Cam follower has spring almost fully retracted; (b) close-up view of impactor at underside of transmitter plate with triggering piezo-film sensors affixed.

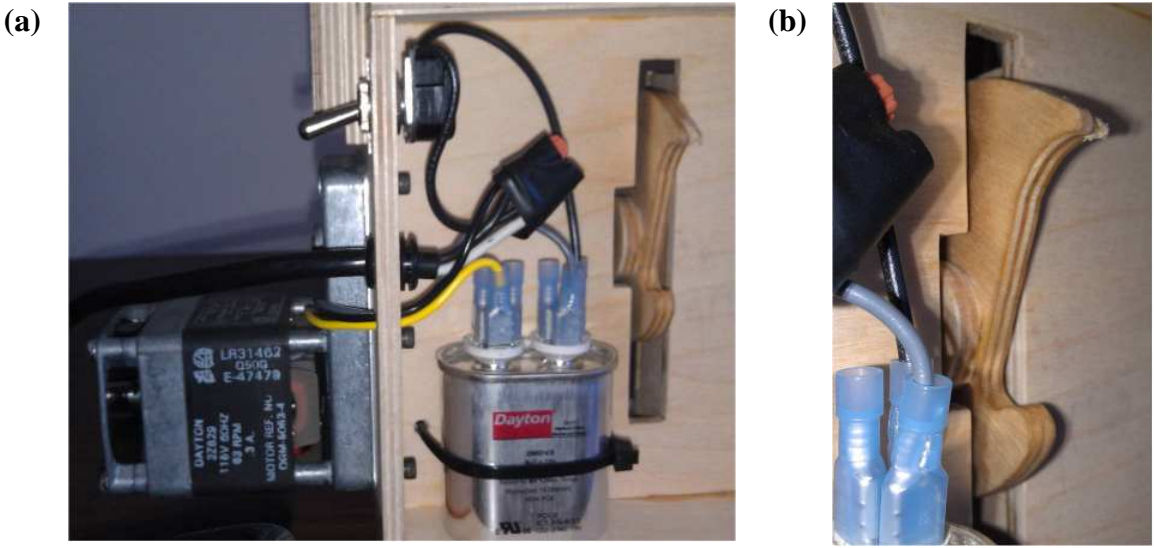

**Figure 8.4. Cam: (a) Mounted in transmitter with motor and starter capacitor; (b) close-up view.**

The rotating cam is driven by the motor, seen in Fig. 8.4(a). Since the motor may be switched on when the spring is nearly fully compressed by the cam follower, the starter capacitor, seen in Fig. 8.4(a), ensures a smooth startup.

The transmitter plate assembly is identical in all respects to the receiver plate assembly, seen in two views in Fig. 8.5. In fact, the two assemblies are interchangeable, except that the leads from the piezo-film sensors affixed to the receiver plate are soldered together in series. The transmitter plate assembly is attached to the body of the transmitter by means of bolts and rubber acoustic isolators, seen in Fig. 8.6. The height of the transmitter plate assembly above the impactor is adjustable to optimize performance.

The L-shaped cutouts at the corners of the transmitter plate assembly, seen in Fig. 8.1 (and in the receiver in Fig. 8.5), allow the position of the transmitter plate to be shifted along both axes parallel to the plate. This capability allows higher-frequency modes of the transmitter plate to be excited by off-center impacts. With four piezo-film sensors arrayed around the center of the plate, we were able to compare the voltage waveforms in each and spectrally analyze the plate resonances one-by-one to determine the nature of the plate normal modes at each resonant frequency.

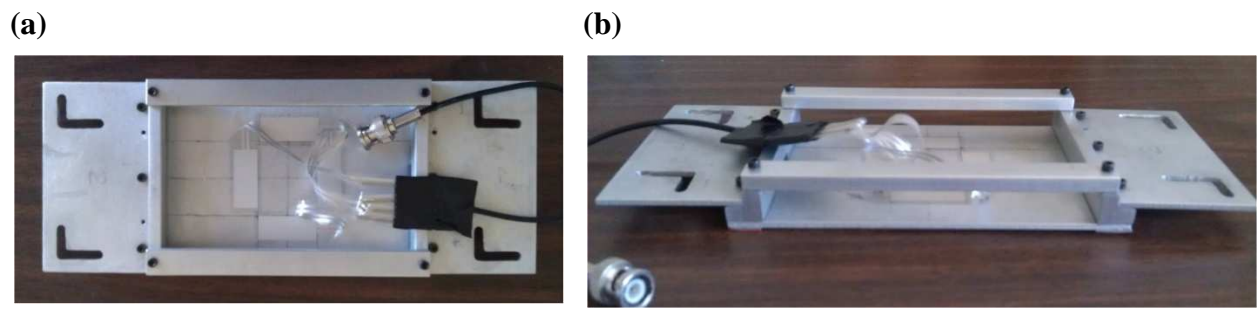

**Figure 8.5. Receiver plate: (a) Underside showing four piezo-film sensors connected in series at a 10-pin header; (b) mounted on wall with 1-mm plate-wall gap.**

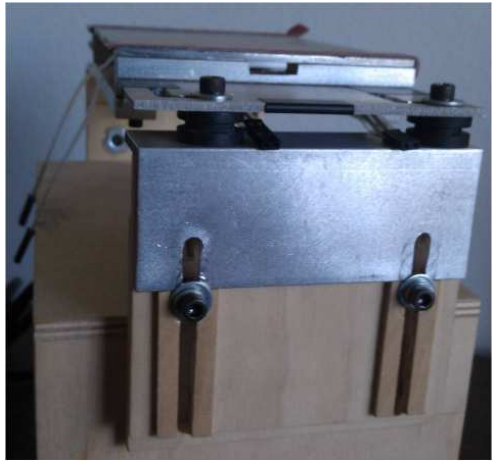

**Figure 8.6. Transmitter plate assembly is attached to body by acoustic isolators. Height of plate above impactor is adjustable to optimize performance.**

The sound pressure level (SPL) of this impact transmitter was not measured. The SPL of an earlier version of the impact transmitter, shown in Fig. 8.7, however, was measured. This earlier version had somewhat lower power than the newer version in Fig. 8.1, because it was not specifically designed to be optimized.

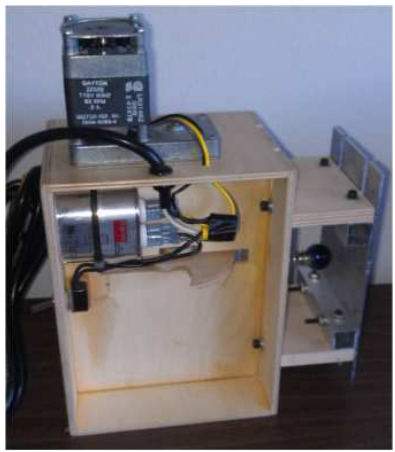

**Figure 8.7. Earlier version of impact transmitter on which SPL measurements were made.**

The equipment used for the calibrated SPL measurements on the earlier transmitter included:
- Bruel & Kjaer (B&K) Model 4189 calibrated microphone
- B&K Model 2669L preamplifier
- B&K Model 4231 94-dB free-field calibrator

All of the following measurements were made in air by averaging over 16 transmitter pulses. All microphone measurements were made by positioning the microphone 1 m directly in front of the transmitter face, with the transmitter resting on a towel for acoustic isolation. No attempts were made to eliminate reverberations and make accurate free-field measurements.

The calibrated root-mean-square (rms) pressure measured during the first 4 ms of the transmitter pulse was $P_{rms}$ = 9.02 Pa. During that 4-ms interval, therefore, the transmitter SPL was 113 dB *re* 20 µPa at 1 m. This SPL is just 7 dB below the threshold of feeling in the ear [28]. At this SPL, the suggested daily noise exposure level for nonoccupational noise is less than 2 minutes [28]. Indeed, to avoid discomfort, it has been necessary to wear ear protection when working with either transmitter, particularly since we were sometimes closer to the transmitter than 1 m.

Figure 8.8(a) is a voltage waveform measured from the earlier transmitter shown in Fig. 8.7. By integrating the PSD of this waveform in Fig. 8.8(b), we find that a 12.5% bandwidth, from 2931 Hz to 3322 Hz, about a central frequency of 3127 Hz, contains 64.1% of all the power of the transmitter from 0 to 50 kHz. The voltage waveform of the transmitter, bandpass-filtered with 12.5% bandwidth from 2931 Hz to 3322 Hz, is shown in Fig. 8.8(c). The rms voltage measured from the first 4 ms of the transmitter waveform in Fig. 8.8(c) is $V_{rms}$ = 71.9 V. Since the amplifier/microphone was calibrated at 9.62 V/Pa, the *in-band* rms pressure during the first 4 ms of the transmitter pulse, therefore, is $P_{rms}$ = 7.47 Pa. During that 4-ms interval, therefore, the *in-band* transmitter SPL in a 12.5% bandwidth was 111 dB *re* 20 µPa at 1 m.

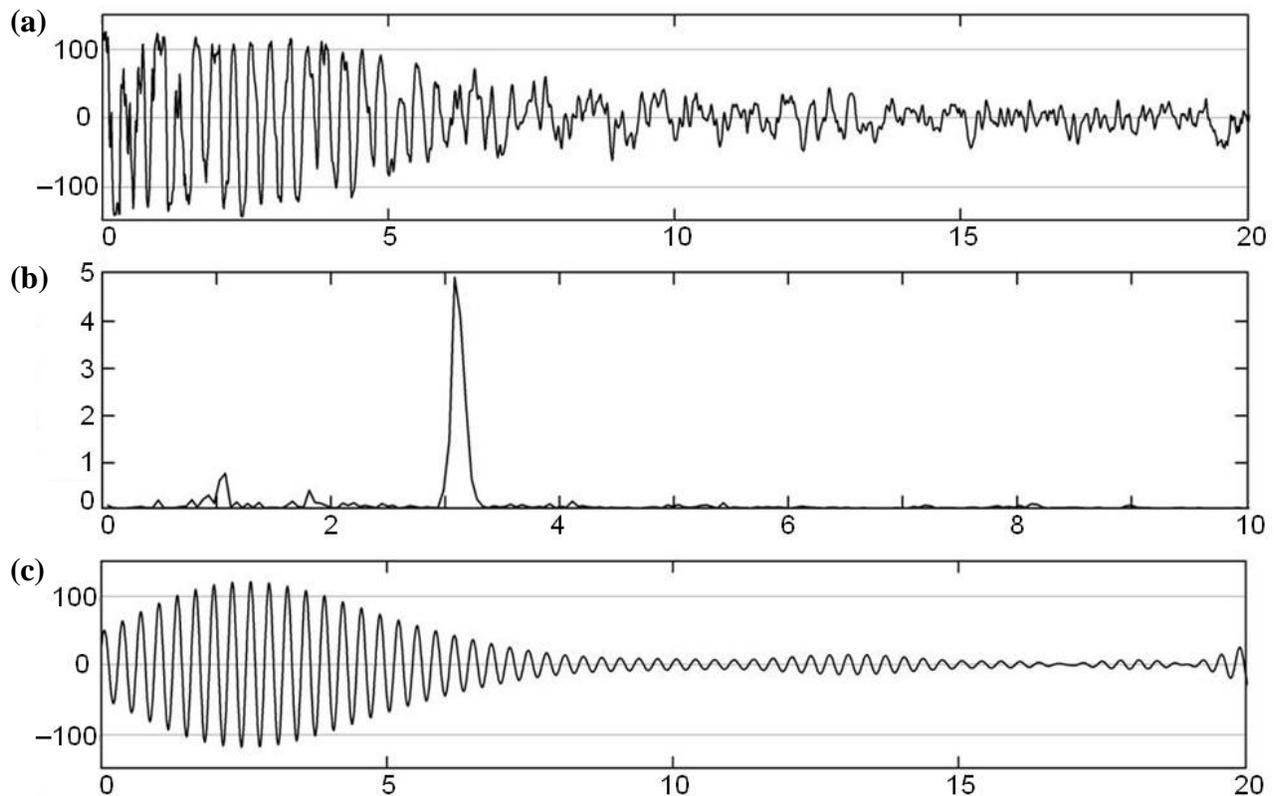

**Figure 8.8.** SPL test of transmitter in Fig. 8.7: (a) Voltage (V) vs. time (ms) calibrated at 9.62 V/Pa; (b) PSD (arb. units) vs. frequency (kHz); (c) Voltage (V) vs. time (ms), as in (a), but bandpass-filtered with 12.5% bandwidth about 3127 Hz.

# 9. Conceptual Design of an Autonomous Sonobuoy Surveillance Network

## 9.1 Long-duration surveillance applications

Owing to ultralow cost and ultralow voltage and power requirements, mechanical impact transducers are potentially ideal candidates for long-duration surveillance missions by proliferated, battery-operated sonobuoys. Specifically, underwater impact transducers like those considered in Sec. 2, with modifications, could be ideal for undersea surveillance by a networked system of fixed autonomous sonobuoy nodes (FASNs) near harbors, bays, and sensitive shoreline installations. As discussed in this section, each battery-operated FASN, like that shown conceptually in Fig. 9.1, could cost less than or about $1000 and, reporting on detection only, have up to a 5-year life. If the concept of self-recharging batteries, presented in Sec. 9.4, is used, the FASN lifetime is not limited by energy storage capabilities. Section 9.5 estimates the range of each FASN for echo-location of submarines in challenging near-harbor noise and shallow water at sea-state 3 to be about 0.25 nmi for the impact-transducer point design of Sec. 2.

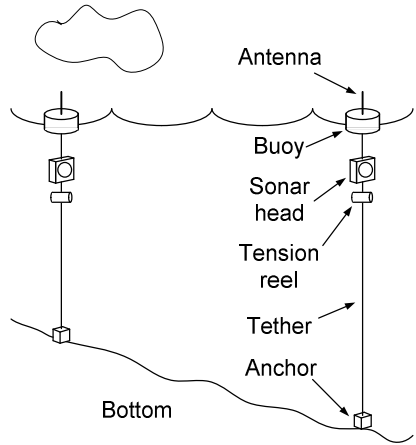

**Figure 9.1. Components of networked long-duration FASNs in littoral waters.**

In the FASN concept, the active-sonar head is buoyed from above and fixed from below by an anchor in the sea bottom. The depth of the sonar head is fixed by the cable length above. A tension reel, acoustically isolated from the sonar head, plays out and retracts the tether below to compensate for varying depth. Of course, the sonar head could also be tethered to an anchor at a fixed distance from the bottom instead, so that the buoy would be submerged and less susceptible to loss.

The batteries, processing, and telemetry subsystems are protected within the sonar head. The buoy supports the communications transponder and transmitter antenna above water. To conserve battery life, the FASN might transmit a communication only upon detection or when the transponder is interrogated for a status update.

Impact transducers with this sonobuoy systems concept should make feasible round-the-clock autonomous submarine surveillance in littoral waters at an extremely low cost per shoreline mile. For years-long deployment in coastal defense, in systems concept and operations concept, sono-buoys using impact-transducer technology can be specifically designed and optimized for networked operation in shallow-water littoral regions and in deep ocean water. A typical wide-area network, protecting the harbors and naval installations around San Diego, California, for example, might involve a grid of about 1600 FASNs, each spaced about 0.5 nmi apart, with the order of 50 or so sonobuoys for receiving and relay functions, together covering about 400 square nautical miles, at an installed cost of the order of $1 million for the sea-based equipment of the surveillance system.

To support multi-year missions of homeland defense and counterterrorism, the transmitters can be fixed in location and depth where they can best surveil and provide early warning for coastal waters near harbors, bays, and other high-value military and civilian targets. The transmitters near the surface are lightweight and do not present a hazard. Also, unlike conventional sonobuoys, the transmitters have little value and are unlikely to be stolen. If they are stolen or disabled, the redundancy in the highly proliferated network will make their loss of negligible consequence to network effectiveness. The source level and other characteristics of the transmitter nodes can be designed to overcome the man-made and ambient noise environments of harbors, to overcome the sea-surface and bottom surface reverberations and volume reverberations in shallow water, and to propagate signals effectively in lossy shallow sea channels.

Conventional submarine surveillance near harbors, bays, and shoreline installations is costly. Sensitive shoreline installations include naval bases, naval air stations, nuclear power plants, and other high-value military, civilian, and government facilities. Submarine surveillance is costly when it is either manpower-intensive or capital-intensive. <u>Passive</u> autonomous sonobuoys can perform submarine surveillance at relatively low cost, but passive sonobuoys have two major problems in shallow-water littoral regions near harbors and shoreline installations.

First, passive sonobuoys, which are basically hydrophone listening devices, have difficulty countering advanced threats, such as quiet submarines and small unmanned undersea vehicles (UUVs), such as the so-called 'narco-subs', which are specifically designed to escape detection by hydrophones. Second, coastal waters near harbors, bays, and high-value installations, exactly where the sonobuoys could be most valuable, are typically extremely noisy environments. Not just the man-made noise of traffic and industrial activity, but the marine life and the turbulence of tidal currents contribute to the noisy ambient environment. By comparison with the Knudsen spectra in deep water, the noise levels in coastal water are 5 to 10 dB higher than in deep water far from shore [31]. In addition, all the commercial active-sonar devices operating in such regions can actually produce spectrum crowding. These noise sources will exacerbate the difficulties of passive sonobuoys in detecting advanced-threat quiet submarines and UUVs.

There is a need for a networked system of low-cost, proliferated units that can be a cost-effective-at-the-margin countermeasure to low-target-strength undersea vehicles in coastal waters. The system should be set-and-forget for years, except for monitoring of the nodes at a network hub. Since the transmitter nodes are active, submerged vehicles cannot escape detection by being quiet. And this systems concept is a cost-effective-at-the-margin responsive countermeasure to UUVs, because the transmitter nodes can be proliferated to as high a number density as needed to detect even the smallest UUVs at a cost much lower than the cost of the UUVs. Conventional manned surveillance, on the other hand, will be no match in cost-effectiveness for proliferated UUVs.

The need for low-cost, round-the-clock surveillance can be fulfilled if certain requirements are met. To avoid the costs associated with a labor-intensive system, the surveillance should be autonomous. For the reasons above, autonomous submarine surveillance may require <u>active</u>-sonar systems. To avoid the costs associated with a capital-intensive system, the equipment should be simple, inexpensive, and have a long operating life with little or no maintenance or replacement needs. Given an average acoustic power output required by a surveillance mission, the battery life of the acoustic source is roughly inversely proportional to its efficiency. Therefore, for a system to be both inexpensive and have a long operating life, unattended, it must be efficient.

Conventional sonobuoys, small sonar sets for undersea listening, which are compact, expendable, and <u>passive</u>, do not seem to be viable means of continuous surveillance for quiet undersea vehicles near harbors and coastal waters. Active sonobuoys based on impact transducers, on the other hand, offer the potential for developing low-cost active sonar technology capable of performing round-the-clock submarine surveillance in the noisy environments of harbors, bays, and coastal waters, unattended, for years.

Conventional transducers for active-sonar sonobuoys, however, cost too much. Conventional means of transduction generally involve high voltages or high currents, which in turn require transformers, amplifiers, and other costly power conditioning components. The power conditioning equipment can, and often does, cost as much as the transducer itself. Sonobuoys and transducer arrays based on conventional technology are not only costly, but also heavy and bulky.

The applicability of impact transducers is expanded, if besides being low-cost, compact, and lightweight, they also have high powers for long ranges, and have high energy-storage capacity and high efficiency or self-recharging capabilities for long-duration operation. Radiated powers of the order of 100 W, such as discussed in Sec. 2, are of interest for wide-area applications, particularly if the low cost allows for profuse proliferation. Total radiation lifetimes of the order of 1000 ping-seconds, corresponding to 100 kJ lifetime radiated energy at 100 W, are of interest for long-duration applications, like years-long surveillance, when used in highly proliferated arrays, for which the dormancy between pulses on each transmitter in a network might be several minutes.

9.2 <u>Anti-submarine warfare (ASW) applications</u>

Impact transducers that are low-cost, compact, and lightweight could potentially open new vistas of applications not just in long-duration surveillance, but also in the more immediate aspects of anti-submarine warfare (ASW) scenarios. The systems concept is to have dumb, but autonomous, transmitters that cost so little and are so compact and lightweight that they can be proliferated quickly and practically over large areas. These transmitters could be dispersed quickly by aircraft or by ships in a wide variety of ASW scenarios, including:
  i)   preparation and conditioning of an ASW battlespace;
  ii)  short-term or long-term surveillance;
  iii) force protection and keep-out maintenance;
  iv)  protection of coastal facilities, such as military bases, harbors, power plants;
  v)   control of waterways and littoral areas.

For example, since conventional ASW sonobuoys are a limited resource, planning their distribu-tion over large littoral areas must be part of engagement planning for tactical ASW.  Ultralow-cost high-power precursor sonobuoys can address the need for environmental assessment of rapidly changing shallow-water battlespaces prior to tactical ASW engagements.  By virtue of their potential for wide proliferation at low cost, these precursor sonobuoys, deployed hours or days before ASW engagements, will allow the most effective and cost-effective deployment and distribution of limited numbers of ASW sonobuoys during tactical ASW.

Since the primary function of environmental surveillance sonobuoys is to conserve and make best use of limited ASW sonobuoy resources, it is important to keep the cost of such sonobuoys small relative to their cost savings.  Impact-transducer technology has a big advantage in this regard, because its effective in-band source levels per dollar will give the surveillance network a tremendous footprint for accurate and reliable environmental measurements.  The cost per square mile for this function and many other ASW functions makes impact transducers a highly attractive alternative.

Many of the challenges of ASW surveillance in littoral waters can be overcome by networks of cheap, proliferated, high-power transmitters.  ASW in littoral waters faces stressing challenges.  In shallow coastal waters, the performance of active acoustic sensors is severely degraded by transmission losses and reverberations.  In shallow water, not only are effects of environmental conditions on acoustic propagation more severe, but littoral environments change quickly over short times and short distances, resulting in dynamically evolving battlespaces.

The function of low-cost transmitters would be to serve as sources of near-isotropic signals, much like beacons.  Echoes from targets would be received for processing at central stations.  The transmitters could serve:  (i) to act with collocated or bistatic receivers to alert the central stations of possible targets of interest in the neighborhood; (ii) to maintain a keepout perimeter by warning off unfriendly forces.

9.3  Homeland security and commercial applications

Potential applications in littoral waters of impact-transducer FASNs that have specifically been cited by the U. S. Department of Homeland Security [30] include:
  i)   wide-area surveillance [for] port and inland waterways, for detection, classification, tracking, and response;
  ii)  persistent, wide-area surveillance … for better maritime border security to … assist in locating illicit activities, materials, or their means of conveyances;
  iii) improved situational awareness by tracking small boat activity, detecting anomalous and/or illegal behavior, and providing timely and actionable information in support of law enforcement and port security efforts;
  iv)  improved detection and tracking of small and large vessels by overcoming environmental clutter issues within the port/harbor as well as in coastal environments;
  v)   protect critical infrastructure and key resources for improved incident response and recovery management along the inland waterways, port/harbor, and coastal regions;
  vi)  more effectively track dangerous cargo being transported on inland rivers and waterways.

Commercial applications for impact-transducer-based autonomous or handheld battery-operated sonar sets include depth sounders, subbottom profilers, fish finding, fisheries aids, divers' aids, position-marking beacons and transponders, and wave-height sensors. Commercial sonar equipment is so widely available for so many applications that spectrum crowding has become a problem in coastal waters. But a powerful, battery-operated active sonar, in either a handheld or a fixed, buoyed configuration, has some ready markets. Handheld configurations will be especially valuable in situations that place a premium on portability, or for which electrical power is limited or unavailable, such as for use by divers for underwater object location. Impact-transducer-based projectors are favored for handheld applications, because they are significantly lighter and smaller than equivalent piezoelectric ceramic marine projectors, particularly when the high-voltage power supplies for the latter are taken into account. The fixed, buoyed configuration might be most valuable when used for position marking as a continuously transmitting beacon or as a transponder that transmits only when interrogated. In either case, the low cost and high efficiency of the device, and the consequent long battery life, are important market advantages.

9.4 Conceptual design of a FASN for long-duration surveillance in littoral waters

As was shown schematically in Fig. 9.1, each FASN unit will be anchored in place by a low-cost weight, like a concrete block, and by a tether. The sonar head will be suspended at either a fixed or a variable depth below a floating buoy. A fixed depth ensures that the water pressure on the thin plate of each impact transducer is roughly constant and the same throughout the FASN network. Then the air or gas pressure behind the plate can be set during manufacture to equalize the water pressure, if necessary. The cable from the sonar head to the buoy will transmit telemetry data to the antenna mounted on the buoy and will transmit control signals from the antenna to the sonar head. A tension reel will remove slack from the tether in changing tides and waves, and will provide a restoring force to keep the buoy positioned nearly vertically above the anchor at all times.

Figure 9.2 shows a schematic diagram of a FASN sonar head that is submersible to variable depths, as for example if each sonar head is to be suspended a fixed distance above the bottom. There is a pressurized cavity behind the resonant plate. The gas pressure in the cavity can be equilibrated to the external water pressure by means of a pressure gauge and a regulator at a valve to a compressed-gas tank.

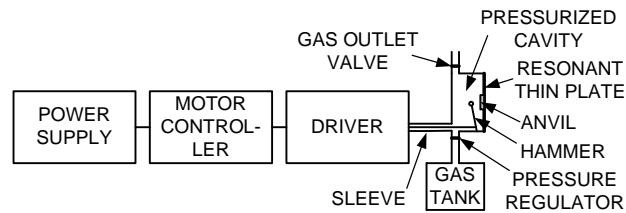

**Figure 9.2. Schematic design of FASN sonar head capable of operating at variable depths.**

The concept for a long-lifetime, autonomous sonar head, described in this section, offers low-cost, efficient, compact, and lightweight approaches to the following basic features: i) energy storage/DC power supply; ii) impact transducer ; iii) watertight body with flotation. A sonar head could in general comprise:
  i)   A watertight pressurized cavity
  ii)  One pair of thin plates

iii) One dual-action actuator for the pair of plates
iv) At least one piezo-film sensor for each plate
v) A self-contained battery power supply
vi) A DSP board and other signal processing and data storage components
vii) A cable or wireless means of transmitting data outside the sonar head

The operating frequency of the impact transducer is chosen mainly as a trade-off between power production and ambient noise. The two principal types of ambient noise at inland rivers and waterways, ports and harbors, and near coastal regions are shipping noise and wind noise. Noise is much higher at low frequencies than high frequencies. In high-noise locations and shallow waters, like the Straits of Malacca, the level of ambient noise from shipping could be 30 dB or more higher at 1 kHz than at 20 kHz [31]. Nevertheless, it is easier to produce acoustic power at lower frequencies because the transducer is bigger and heavier. For this conceptual design example, 1 kHz is chosen as the nominal operating frequency. And for this example, the point design of a 1-kHz underwater impact transducer from Sec. 2 is used.

For the point design, the energy estimates of Table 2.3 showed that a battery energy/pulse of 3.3 J was needed for an impact transducer to radiate a 1-kHz, 4-ping-ms pulse at an average power of 50W. A single high-capacity lithium thionyl chloride battery, such as the Saft LS 33600 3.6-V "D-cell" size battery, has a 17-A·hr capacity. This capacity corresponds to 220 kJ of stored energy per battery, enough to produce 67,000 pulses, for a 270 ping-second lifetime at 50 W average power. At a prf of one pulse per 5 minutes, one battery could drive the impact transmitter of the point design for nearly 8 months, and 8 such "D-cell" size batteries, for 5 years.

An alternative power supply for a FASN comprises one or two rechargeable batteries and a self-recharging generator. The self-recharging generator concept is the following: As the tension reel attached to the tether (shown in Fig. 9.1) unwinds and rewinds with wave motion, the spinning reel drives an electrical generator, which recharges the batteries. Through a ratchet-like cable drive, the wave motion acting on the buoy can be made to do work on the generator, and to produce and store electrical energy in the rechargeable batteries, whether the tension reel is unwinding or rewinding. The cable should be made of a strong, lightweight material that does not stretch significantly under tension and that has a specific gravity less than or about equal to that of sea water, 1.03. That will keep the requirements for tension in the reel and buoyancy in the buoy modest, even in deep littoral waters. Unlike intermittent and unreliable solar power, wave power is virtually continuous. The minimal average-power requirements of each FASN, including transmit/receive functions, 2-way communications, and station-keeping, can easily be provided by a self-recharging generator such as this one, based on wave motion.

The sonar projector can be designed to produce two (or more) counter-directed pulses, one each from two opposite faces of the sonar head. The full-angle diffraction-limited divergence of a beam of wavelength $\lambda$ from a baffled circular piston of radius $a$ is $\theta_{DL} = 1.22\lambda/a$. For the 1-kHz point design ($\lambda = 1.5\,\text{m}$), the beam divergence is much greater than 1 radian, and each beam will cover at least a $2\pi$-steradian half space nearly isotropically. Both beams together will provide $4\pi$ coverage.

A dual-plate impact-transmitter design, like that in Fig. 9.3, has a further advantage of allowing identically manufactured FASN units to produce unique pulse signatures, even though they operate at identical frequencies, bandwidths, and pulse widths. As discussed in Sec. 7, impact transmitters have a sharply defined pulse start time, $t = 0$. Piezo-film sensors affixed to the plate register a sharply defined voltage spike at the moment of impact of the impactor on the plate. The jitter $\Delta t$ in determining $t = 0$ is very much less than a wave period, and is generally shorter than the voltage sampling time. The high precision to which $t = 0$ is known can be used in the following way to assign a unique pulse signature to each dual-plate FASN sonar head after it has been manufactured.

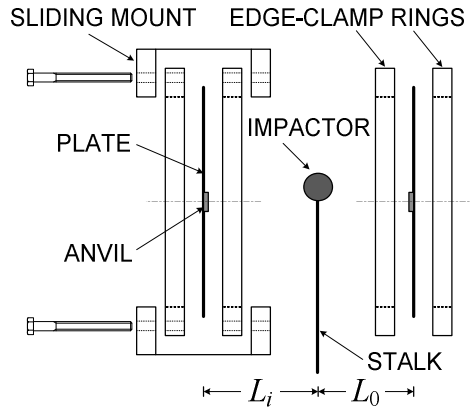

**Figure 9.3. Dual-plate impact-transmitter with adjustable impactor-plate offset $L_i$ for unique signature (exploded view).**

Suppose the baseline design is a dual-plate impact transmitter, like that shown in Fig. 9.3, in which the second plate is struck by the same impactor immediately after the first. That is, the driver mechanism causes the impactor after it strikes the first plate to accelerate towards the second plate. The offset distance of the impactor to the second plate, shown as $L_i$ in Fig. 9.3, is adjustable. Either the plate position can be adjusted by means of a sliding mount, as shown in Fig. 9.3, or the impactor position between two fixed plates can be adjusted. When $L_i$ is changed, the difference in the pulse start times of the two plates is changed. Since the pulse start time at each plate is known to high precision, the difference in the times is also known precisely and can be used to identify the source.

Each thin plate could have a steel button, called an anvil, affixed to its center. The anvil has several purposes. It protects the thin plate from hundreds of thousands of impacts of the hammer to be expected during the lifetime of the plate. (In Fig. 9.2, the impactor and stalk together are referred to as the 'hammer'.) It produces a more elastic collision with the impactor, which may also be steel. And the anvil may help to put more of the impactor kinetic energy into exciting the desired mode of oscillation and less into parasitic modes, if the hammer strikes slightly off-center.

For the baseline design of a network of hundreds of 1-kHz FASNs, suppose the sampling rate is 32 samples per ms. The difference in start times of the pulses from each of the two faces, therefore, is known to within about 30 μs. Suppose the offset $L_i$ is adjustable over about 1 cm. In the baseline design of Sec. 2, the velocity of the impactor upon impact is of the order of 1 m/s. That means the difference in pulse start times at the two plates can vary over about 10 ms as $L_i$ is adjusted over the 1-cm adjustable range. In this example, of the order of 300 uniquely distinguishable offset positions of the second plate are available over the 1-cm adjustable range. The adjustments on the FASNs can be made either by dialing in random offsets $L_i$ or by fine positioning by means of a micro-positioning knob, similar to a micrometer, controlling a worm-screw drive.

Upon producing a double pulse, a FASN will transmit either to a relay buoy or to a central station a signal conveying its location and the time difference between the start of its two pulses. Any FASN receiver that receives this double pulse, and then a short time later receives the reflected double pulse, will transmit a signal conveying its location and the times of receiving the pairs of double pulses. From this information, the range of the target to the receiver is known. If at least two FASNs in the network report a detection, then the position of the target can be determined by triangulation.

This method of identifying the source depends upon both pulses from a single FASN reflecting from a target and being received at any other FASN in the network. That means the method depends upon the impact transmitters being essentially simple sources with nearly isotropic radiation patterns from each plate.

9.5 FASN network for ASW in harbor noise

In a FASN network, the primary means of mitigating risk in the range of each node is by proliferating nodes to achieve the desired margin of safety. To determine the sensitivity of range of each FASN to operational parameters, and to ensure that the conceptual design in Sec. 9.4 can be extrapolated practically to a FASN product, we have evaluated the active sonar equation for shallow-water operation.

The active-sonar equation for the monostatic case, in which the transmitter and receiver are collocated, is given by [31]

$$TL = \tfrac{1}{2}(SL + TS + DI - NL - DT),  \qquad (9.1)$$

where $TL$ is one-way transmission loss, $SL$ is transmitter source level, $TS$ is target strength, $DI$ is receiving directivity index, $NL$ is level of ambient noise and self-noise, and $DT$ is the detection threshold of the receiver. Transmitting directivity index, or gain, is included in $SL$. Since the sonar equation is used here to estimate the maximum range for detection by a FASN in noisy environments, we will assume that the range is long enough to be noise-limited rather than reverberation limited. Then the background level $NL$ is isotropic noise, rather than reverberation.

Using this active-sonar equation for the monostatic case gives a conservative estimate of the range, because in a network of FASNs, the echo from the target may be received by a FASN that is closer than the transmitter, in which case the total transmission loss will be less than the $2TL$ used in Eq. (9.1).

In our preliminary concept based on the point design of an impact transmitter in Sec. 2, FASN operation has the following characteristics: Average pulse power $\bar{P}$ is 50 W; pulse duration $T$ is 4 ms; transmitting (and receiving) directivity index or gain $G$ is 0 dB; probability of false alarm $P_{FA}$ is 1%; probability of detection $P_D$ is 50%; frequency is 1 kHz; and bandwidth $B$ is 91 Hz. The false alarm and detection requirements for each individual FASN are modest, because the probability of detection by a network of hundreds of FASNs will be much closer to 100%, and the network probability of false alarm will be much lower than 1%.

The echo pulse will be much longer than the 4-ms transmitter pulse, reducing the effective source level of the transmitter. The typical echo duration produced by a submarine target with multipath propagation in shallow water is 100 ms [31]. The effective source level is therefore reduced by about $10\log[(4\,\text{ms})/(100\,\text{ms})] = -14$ dB [31]. But because this concept is for a double-pulse, dual-plate sonar, with effectively twice the power or twice the pulse duration, the effective source level is increased by 3 dB. Then the source level of a 50-W FASN transmitter at 1 yd, relative to 1 µPa, is

$$SL = 171.5 + 10 \log(\overline{P}) - 14 + 3 + G = 177.5 \text{ dB}. \qquad (9.2)$$

The nominal target strength of a submarine at beam aspect is +25 dB, at bow or stern aspect is +10 dB, and at intermediate aspect is +15 dB [31, 32]. For our estimates, we choose the nominal intermediate-aspect submarine target strength, $TS = +15$ dB. This estimate is conservative, since surveillance by a network of fixed sonobuoys will generally allow targets approaching coastal waters to be viewed from a wide range of aspect angles.

Figure 9.4 shows ambient noise levels in bays and harbors from World War II data. For average noise locations, the noise level at 1 kHz is about 65 dB. To protect entrances to bays and harbors, a FASN network can stand off some distance from the noisiest locations, such as the entrances, for an improvement of say 5 dB to $NL = 60$ dB at 1 kHz.

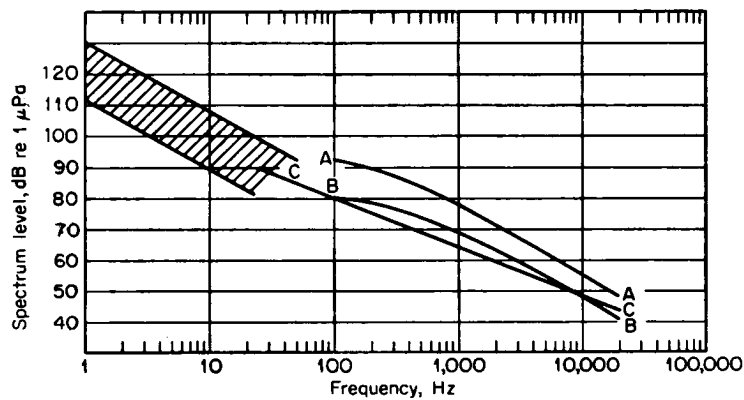

**Figure 9.4.** Noise levels (dB re 1 µPa) in bays and harbors vs. frequency (Hz): A) High-noise location: entrance to New York Harbor in daytime; B) Average-noise location: upper Long Island Sound; C) Average of many WW II measurements. Adapted from [33].

The detection threshold $DT$ of a FASN receiver is relatively low because the signal, shown for example in Fig. 2.3, is known. The optimum receiver for a known signal is a cross-correlator, which correlates the known signal with the measured signal plus noise background. For the case of a known signal, the detection threshold becomes [31]

$$DT = 10\log(S/N_1) = 10\log(d_i/2T) = 29 \text{ dB}, \tag{9.3}$$

corresponding to the input signal-to-noise ratio referred to a 1-Hz band of noise, $S/N_1 = 750$. Receiver-operating-characteristic (ROC) curves of $P_D$ vs. $P_{FA}$, such as given in [31], were used to determine the detection index, $d_i = 6$, for $P_D = 50\%$ and $P_{FA} = 1\%$.

From Eq. (9.1), therefore, we find the allowed one-way transmission loss of a FASN system at the submarine detection threshold in harbor noise for $P_D = 50\%$ and $P_{FA} = 1\%$ is

$$TL = \tfrac{1}{2}(SL + TS + DI - NL - DT) = \tfrac{1}{2}(177.5 + 15 + 0 - 60 - 29) = 52 \text{ dB}. \tag{9.4}$$

The semi-empirical expression for transmission loss in shallow seawater at short ranges, conditions amply satisfied by all FASN units in littoral waters, is [34]

$$TL = 60 + 20\log(R) + \alpha R - k_L, \tag{9.5}$$

where $R$ is the range in kiloyards, $\alpha$ is the absorption coefficient of seawater, and $k_L$ is the near-field anomaly, which is dependent on sea-state and bottom type. At 1 kHz, $\alpha$ is negligible, about 0.07 dB/kyd [31], and $k_L$ ranges from about 3 dB at sea-state 1 to 2 dB at sea-state 5, for both sand and mud bottoms. Solving Eqs. (9.4) and (9.5) for an intermediate value of $k_L = 2.5$ dB gives a maximum range for detection of submarines in littoral waters and harbor noise of about $R = 520$ yd. Of course, the detection range will be substantially less in high winds and heavy rains. But this range estimate of about 0.25 nmi shows that an ultralow-cost, 1-kHz, 50-W, 200-mJ/pulse impact transducer can potentially perform useful submarine surveillance in near-harbor noise, in shallow water at sea-state 3, with a littoral network of fixed nodes. Depending on distribution, the FASNs could be spaced as much as 0.5 nmi apart. Each battery-operated FASN could have up to a 5-year life with 8 "D-cell" size batteries or, if self-recharging, have a lifetime not limited by energy storage constraints.

The total mass of a dual-plate, 1-kHz sonar head is less than or about 10 kg. Most of that mass (> 5.4 kg) is in the two aluminum plates and the steel impactor, as shown in Sec. 2. The entire sonar head can be packaged in a volume of about 1 cu. ft. A 10-kHz sonar head, on the other hand, could be built much smaller and lighter, primarily because the impact-transducer plates are much smaller and thinner and the impactor is much lighter at 10 kHz. The trade-off, however, is that the impact transmitter produces less power at 10 kHz. For comparison with the range estimate above for a 1-kHz sonar, the following is the range estimate for a 10-kHz system with the same 4-ms ping and an average power of $\bar{P}_{10}$. Without calculating a point design at 10 kHz, we will simply estimate that the 15-cm wavelength under water is sufficiently greater than the active radius of the thin plates that the radiation is effectively isotropic, and there is no directional gain.

Then from Eq. (9.2), the source level of a 10-kHz transmitter is $SL = 160.5 + 10\log(\bar{P}_{10})$. From Fig. 9.4, the average harbor noise level at 10 kHz is about 48 dB. With some stand off from harbor entrances, a 5-dB improvement would give $NL = 43$ dB. The detection index and detection threshold are the same, $P_D = 50\%$ and $P_{FA} = 1\%$. Then the allowed one-way transmission loss of a 10-kHz FASN system at the submarine detection threshold in near-harbor noise, for $P_D = 50\%$ and $P_{FA} = 1\%$, is $TL = 5\log(\overline{P}_{10}) + 54$ dB.

The absorption coefficient at 10 kHz is about 0.9 dB/kyd [31], so we estimate that $\alpha R = 1$ dB. Then for $k_L = 2.5$ dB as before, the transmission loss in shallow seawater at short ranges, from Eq. (9.5), is $TL = 58.5 + 20\log(R)$. Equating these two expressions for transmission loss at 10 kHz gives the range of a 10-kHz impact-transducer sonar in harbor noise as

$$R = (610 \text{ yd})[\overline{P}_{10}(\text{W})]^{1/4}. \tag{9.6}$$

Even if the transmitter produces only 1 W of average power in a 4-ms ping, the range of the 10-kHz transmitter at 1 W is comparable to the range of the 1-kHz transmitter at 50 W. The reason is primarily that average harbor noise is about 17 dB quieter at 10 kHz than at 1 kHz. In littoral areas where ambient noise is not as significant, a 50-W 1-kHz transmitter would have much greater range than a 1-W 10-kHz transmitter.

Not only is a 10-kHz impact-transducer sonar head much lighter and more compact, it is also less costly than a 1-kHz sonar head of less than or about $1000. The power requirements of the 10-kHz system are lower and, to the extent its range may be more limited, shortcomings in range can be compensated by greater proliferation of FASNs in the network. As a general rule, however, noisy environments favor higher frequencies.

## 10. Summary and Conclusions

This paper presented a new concept for low-cost, high-power mechanical acoustic transmitters and matched resonant receivers. A mechanical impact transmitter converts stored mechanical energy into kinetic energy of a mass, which strikes a thin plate. The plate vibrates from the impact, primarily in the resonant modes of vibration of the plate, and the vibrations drive sound waves into the medium on the other side of the transmitter plate at the resonant frequencies of the plate. While a diaphragm is much more efficient than a plate at generating sound waves in air, a thin plate can be designed to be highly efficient at generating sound in dense media.

Impact transmitters are capable of producing acoustic pulses at one or more discrete resonant frequencies with much greater output power than input power, by storing mechanical energy over times much longer than the pulse widths. This power compression allows the transducers to be powered by COTS batteries at volts or tens of volts, instead of expensive, heavy and bulky power modulation equipment operating at kilovolts.

These mechanical/impact transducers are well suited for coupling acoustic pulse energy into dense media, such as walls and water. This paper discussed two such applications: Through-wall surveillance (TWS), that is, the detection and tracking of persons through walls; and years-long, maintenance-free underwater surveillance by a low-cost network of autonomous, battery-operated sonobuoys.

Acoustic TWS has two major advantages over wideband radar: Acoustic TWS systems can detect and track humans through metal walls, such as the walls of trailer-truck bodies and cargo containers; and acoustic TWS, operating in narrow frequency bands at short wavelengths, can detect even stationary persons by their millimeter-scale breathing, while wideband radar is limited to detecting velocities greater than about 10 cm/s.

For sound to penetrate a wall, reflect from objects and persons on the other side, and return to a receiver, it must pass through at least four wall/air interfaces. Since orders of magnitude of power are lost at each interface, efficient coupling of acoustic power from transmitters into walls and from walls into receivers is essential to successful TWS. Section 4 showed that efficient coupling of an impact transmitter to a wall can be achieved by effectively turning the wall into a soundboard for the transmitter. The 'bridge' between the transmitter and the wall is a set of spacers that do not interfere with the oscillations of the transmitter plate, but that maintain a thin air gap – the thinner the better – between the plate and the wall. The power transmitted into the wall is inversely proportional to the square of the air gap. A similar 'bridge' efficiently couples acoustic energy from the wall into the receiver.

If the resonant transmitter and receiver plates are not one and the same, then narrowband operation of the TWS system is achieved by exactly matching the transmitter plate and its boundary conditions and resonances to the receiver plate and its boundary conditions and resonances. Piezoelectric film sensors on the transmitter plate can serve to establish the exact moment of impact and the beginning of the acoustic pulse. Multiple small piezoelectric film sensors on the receiver plate, positioned where the modes of oscillation produce the greatest stress, and connected in series, produce the maximum signal strength for a given oscillation amplitude. If $n$ sensors connected in series receive equal signals, the signal strength is amplified by $n^2$.

Section 7 demonstrated detection through a massive wooden desk and a solid wooden table using the breadboard impact transmitter and matched resonant receiver described in Sec. 8. Since no real-time signal processing was available for managing and comparing multiple waveforms, the detection of a moving person was demonstrated in post-processing, in Figs. 7.3 and 7.7. Section 7 showed that a narrowband filter/amplifier could have increased S/N of the breadboard system up to 26 dB, by excluding noise outside the resonant bandwidth. But Figs. 7.4 and 7.5 also showed that if the bandpass filter is too narrow, probability of detection can be degraded. To avoid such degradation, the bandpass filter should be much wider than the inverse of the record length.

Section 6 demonstrated an algorithm that was used to track and display the locations of multiple persons behind a wall in near real time with a resolution less than or about 1 foot. The tracking algorithm requires signals from at least two receivers as inputs for triangulation. An array of lightweight, low-cost, battery-operated TWS transducers can provide high probability of detection through walls over large areas.

Section 9 presented a conceptual design of a low-cost, long-duration sonobuoy surveillance net-work for harbors and littoral waters. The sonobuoy units, or nodes, of the network may be designed for specific missions based on the projector design equations presented in Sec. 2. Equation (2.15) is a set of more than 20 interrelated design equations, parametrized in terms of just a few features. These design equations are highly nonlinear because impact transmitters operate most efficiently underwater when the reactive mass of the water outside the plate is comparable to the active mass of the plate.

Because each sonobuoy node in a network is based on a simple, lightweight sonar comprising a mechanical impact transmitter and matched resonant receiver, operating at low voltage on COTS battery power, these nodes can be profusely proliferated in dense networks at low cost per surveilled coastline mile. Section 9 used the shallow-water sonar range equation to determine the node density required to achieve specified levels of probability of detection and probability of false alarm.

A fully cooperative network can achieve surveillance synergies. For example, a target cannot slip through a network without exposing its aspect angle of highest target strength to at least some of the sonobuoys some of the time. Probability of network detection degrades gracefully with depletion of nodes. Echoes from a target of a pulse produced by a distant sonobuoy can be detected by a sonobuoy closer to the target, effectively reducing the required range of each sonobuoy and increasing the probability of network detection. Section 9 discloses a novel method for uniquely identifying the source of a pulse from its detected echo in a highly proliferated network, even when each of the sonobuoys is manufactured identically. Section 9 also discloses a novel concept for self-recharging sonobuoys in shallow waters, so that maintenance and replacement schedules do not need to be determined by battery numbers and capacities.

In summary, the low cost and high-power capabilities of mechanical impact transducers make them attractive candidates for applications requiring the repetitive production of narrowband acoustic pulses in dense media, such as walls and water.

### Appendix A. Abstract of U. S. Patent Application, "High-power mechanical transducers for acoustic waves in dense media" [35]

Mechanical transducers efficiently produce and couple high-power acoustic pulses into liquid and solid media. In a transmitter, mechanical excitation of a thin transmitting plate is provided by an actuator or a motor that causes a mass to strike or a drive rod to push the thin plate. If struck, as the thin plate rings down, it delivers much of its kinetic energy to acoustic radiation in the dense medium. Different mechanisms may be used to excite the plate, and different mechanisms may be used to couple the plate excitations into dense media. Conditions are found for efficient transduction of mechanical energy by a thin plate into acoustic radiation in solid and liquid media. A receiver comprises a plate having matching resonances to the transmitting plate. Discrete narrowband frequencies of acoustic signals are used to detect phase changes in waves reflected from a moving object. By interfering successive return pulses, small changes in phase and amplitude within the reflected beam lead to large changes in interfering voltage waveforms.

**References**

1. F. S. Felber, H. T. Davis III, C. Mallon, and N. C. Wild, "Fusion of radar and ultrasound sensors for concealed weapons detection," in *Signal Processing, Sensor Fusion, and Target Recognition V*, I. Kadar, V. Libby, Eds., Proc. SPIE 2755, 514 – 521 (1996).

2. F. S. Felber, C. Mallon, N. C. Wild, and C. M. Parry, "Ultrasound sensor for remote imaging of concealed weapons," in *Command, Control, Communications, and Intelligence Systems for Law Enforcement*, Edward M. Carapezza, Donald Spector, Eds., Proc. SPIE 2938, 110 – 119 (1997).

3. F. Felber, N. Wild, S. Nunan, D. Breuner, and F. Doft, "Handheld ultrasound concealed weapons detector," in *Enforcement and Security Technologies*, A. Trent DePersia, John J. Pennella, Eds., Proc. SPIE 3575, 89 – 98 (1998).

4. F. Felber, "Concealed weapons detection technologies," Final Technical Report AFRL-IF-RS-TR-1998-218 (Dec. 1998); http://www.dtic.mil/dtic/tr/fulltext/u2/a359517.pdf

5. F. Felber *et al*., "Handheld remote concealed weapons detector," Final Technical Report, Nat. Inst. Justice Doc. 178564 (Feb. 1999); https://www.ncjrs.gov/pdffiles1/nij/grants/178564.pdf

6. N. C. Wild, F. Doft, D. Breuner, and F. S. Felber, "Handheld ultrasonic concealed weapon detector," in *Technologies for Law Enforcement*, Edward M. Carapezza, Ed., Proc. SPIE 4232, Paper 4232-27 (2000).

7. D. D. Ferris and N. C. Currie, "Microwave and millimeter wave systems for wall penetration," *SPIE Conference on Targets and Backgrounds: Characterization and Representation IV*, Orlando, FL, April 1998, Proc. SPIE 3375, 269 – 279 (1998).

8. United Airlines High Street Emporium, "This barking dog alarm works through doors and walls," p. 79 (Summer, 1995).

9. F. Su, "Surveillance through walls and other opaque materials," OE Reports, No. 140, p. 1, SPIE (August 1995).

10. L. M. Frazier, "Surveillance through walls and other opaque materials," Proc. IEEE Nat. Radar Conf. (Ann Arbor, MI, 13–16 May 1996).

11. J. A. Lovberg, J. A. Galliano, S. E. Clark, "Passive millimeter-wave imaging for concealed article detection," in *Command, Control, Communications, and Intelligence Systems for Law Enforcement*, Edward M. Carapezza, Donald Spector, Eds., Proc. SPIE 2938, 120 – 130 (1997).

12. L. M. Frazier, "Radar surveillance through solid materials," in *Command, Control, Communications, and Intelligence Systems for Law Enforcement*, Edward M. Carapezza, Donald Spector, Eds., Proc. SPIE 2938, 139 – 146 (1997).

13. D. G. Falconer, K. N. Steadman, D. G. Watters, "Through-the-wall differential radar," in *Command, Control, Communications, and Intelligence Systems for Law Enforcement*, Edward
52

M. Carapezza, Donald Spector, Eds., Proc. SPIE 2938, 147 – 151 (1997).

14. G. R. Huguenin, "Millimeter-wave concealed weapons detection and through-the-wall imaging systems," in *Command, Control, Communications, and Intelligence Systems for Law Enforcement*, Edward M. Carapezza, Donald Spector, Eds., Proc. SPIE 2938, 152 – 159 (1997).

15. Greneker, Eugene F., "Radar flashlight for through-the-wall detection of humans," *SPIE Conference on Targets and Backgrounds: Characterization and Representation IV*, Orlando, FL, April 1998, Proc. SPIE 3375, 280 – 285 (1998).

16. L. M. Frazier, "Surveillance through nonmetallic walls," in Enforcement and Security Technologies, A. Trent DePersia, John J. Pennella, Eds., Proc. SPIE 3575, 108 – 112 (1998).

17. Time Domain Corp., http://www.timedomain.com/news/wall.php (April 24, 2006).

18. L. W. Fullerton, U.S. Patent 7,030,806, "Time domain radio transmission system," issued April 18, 2006.

19. I. Cowie, "Through-wall surveillance for locating individuals within buildings," Final Technical Report, Nat. Inst. Justice Doc. 228898 (Feb. 2010); https://www.ncjrs.gov/pdffiles1/nij/grants/228898.pdf

20. Lawrence Livermore National Laboratory, "Urban Eyes: Ultra-wideband (UWB) detecting and tracking," Presentation UCRL-PRES-217242; https://ipo.llnl.gov/data/assets/docs/UrbanEyesPresentation.pdf

21. V. Orphan, "Advanced cargo container scanning technology development," MTS R&T Coordination Conf. (Wash., DC, 2004); http://onlinepubs.trb.org/onlinepubs/archive/Conferences/MTS/3A%20Orphan%20Paper.pdf

22. http://en.wikipedia.org/wiki/File:VACIS_Gamma-ray_Image_with_stowaways.GIF

23. N. C. Wild, F. S. Felber, M. Treadaway, F. Doft, D. Breuner, and S. Lutjens, "Ultrasonic through-the-wall surveillance system," in *Technologies for Law Enforcement*, Edward M. Carapezza, Ed., Proc. SPIE 4232, Paper 4232-29 (2000).

24. C. H. Sherman and J. L. Butler, Transducers and arrays for underwater sound (Springer, NY, 2007).

25. F. Felber, "High-Power Stealthy Acoustic Through-the-Wall Sensor," Starmark Final Report No. M67854-02-C-1010 (Marine Corps Systems Command, Quantico, VA, 16 May 2002).

26. J. Delany, "Bender transducer design and operation," J. Acoust. Soc. Am. 109(2), 554-562 (2001).

27. J. Crawford, C. Purcell, and B. Armstrong, "A modular projector system: Modeled versus measured performance," UDT Europe 2006 (2006).

28. L. E. Kinsler, A. R. Frey, A. B. Coppens, and J. V. Sanders, <u>Fundamentals of Acoustics</u>, 3rd Ed. (Wiley & Sons, NY, 1982).

29. M. Sansalone and W. B. Streett, <u>Impact-Echo: Nondestructive Testing of Concrete and Masonry</u> (Bullbrier Press, Jersey Shore, PA, 1997).

30. U. S. Dept. Homeland Security, Science & Technology Long-Range Broad Agency Announcement 11-03 (24 Jan. 2011).

31. R. J. Urick, <u>Principles of Underwater Sound</u>, 3rd Ed. (McGraw-Hill, NY, 1983), p. 213.

32. <u>Principles and Applications of Underwater Sound</u> (Dept. Navy, HQ Naval Material Command, Wash., DC, 1968), p. 165.

33. V. O. Knudsen, R. S. Alford, and J. W. Emling, J. Mar. Res. <u>7</u>, 410 (1948).

34. H. W. Marsh and M. Schulkin, J. Acoust. Soc. Am. <u>34</u>, 863 (1962).

35. F. S. Felber, U. S. Patent Application 14/250,223, "High-power mechanical transducers for acoustic waves in dense media," filed April 10, 2014.